\documentclass[pre,twocolumn,footinbib,floatfix,twoside]{revtex4}
\usepackage{amsmath,bbm,graphicx,bm}

\newcommand{\deriv}[2]{\frac{\partial #1}{\partial #2}}

\begin{document}

\title{Discontinuous Molecular Dynamics for Rigid Bodies: Applications}

\author{Lisandro Hern\'{a}ndez de la Pe\~{n}a, Ramses van Zon, 
        Jeremy Schofield} 

\affiliation{Chemical Physics Theory Group, Department of Chemistry,
             University of Toronto, Ontario, Canada M5S 3H6}

\author{Sheldon B. Opps} 

\affiliation{Department of Physics, University of Prince Edward Island,
         550 University Avenue, Charlottetown, Prince Edwards Island,
         Canada C1A 4P3}
\date{July 20, 2006}

\begin{abstract}
Event\mbox-driven molecular dynamics simulations are carried out on
two rigid body systems which differ in the symmetry of the{ir}
molecular mass distribution{s}.  First, simulations of methane in
which the molecules interact via discontinuous potentials are compared
with simulations in which the molecules interact through standard
continuous Lennard\mbox-Jones potentials. It is shown that under
similar conditions of temperature and pressure, the rigid
discontinuous molecular dynamics method reproduces the essential
dynamical and structural features found in continuous\mbox-potential
simulations at both gas and liquid densities.  Moreover, the
discontinuous molecular dynamics approach is demonstrated to be
between 2 to 100 times more efficient than the standard molecular
dynamics method depending on the specific conditions of the
simulation.  The rigid discontinuous molecular dynamics method
is also applied to a discontinuous\mbox-potential model of a liquid
composed of rigid benzene molecules, and equilibrium and dynamical
properties are shown to be in qualitative agreement with more detailed
continuous\mbox-potential models of benzene. Qualitative differences
in the dynamics of the two models are related to the relatively crude
treatment of variations in the repulsive interactions as one benzene
molecule rotates by another.
\end{abstract}

\maketitle

\section{Introduction}

Although event\mbox-driven simulation studies of systems of
molecules interacting via discontinuous potentials are
common\cite{DMDref1,DMDref2,DMDref3,Chapelaetal84,Chapelaetal89,DMDref4},
most work to date has considered fully\mbox-flexible models of
molecular systems in which standard ``bonded'' potentials for
the motion of bonds, bond angles, and internal dihedral angles
are replaced by stepped, discontinuous versions of these
interactions\cite{karplus_1999}.  However{,} much of the internal
motion of a molecule is largely unimportant in understanding many
dynamical properties of condensed phase systems on long time scales.
As the efficiency of an event\mbox-driven simulation depends
critically on the number of events to process per unit of simulation
time, a great deal of computational effort in simulations of large
molecular systems is devoted to treating essentially
unimportant internal motion of molecules, potentially limiting the
scope of the simulation.  Similar issues are relevant in more
standard, continuous\mbox-potential models of molecular systems.  In
this context, methods of incorporating partial internal constraints in
the dynamics of molecules for performing efficient simulations of
rigid systems with quaternions and matrix propagation schemes have
been developed to extend the time scale or system size accessible to
simulation.

In the preceding paper\cite{preceeding}, a general framework to carry
out event\mbox-driven simulations of molecular systems interacting via
discontinuous potentials in the presence of constraints was outlined.
This formalism is generally applicable to both semi\mbox-flexible and
fully\mbox-rigid bodies.  In the case of a rigid molecular system, it was
shown that the analytical solution of the free motion of arbitrary
rigid bodies can be combined with efficient numerical algorithms for
the search for event times and the use of appropriate collision rules
to carry out event\mbox-driven or discontinuous molecular dynamics (DMD)
simulations.

In the present work, we apply this approach to two molecular systems
of practical interest, methane and benzene.  The lack of similarity in
the mass distribution of methane and benzene molecules, classified as
spherical and symmetric top rotors, respectively, implies that the
rotational motion in these two systems is quite different.  First, a
discontinuous\mbox-potential model of rigid molecular methane is
constructed by analogy with a standard continuous\mbox-potential
model based on Lennard\mbox-Jones interactions.  Structural and
dynamical characteristics of DMD simulations obtained at gas and
liquid\mbox-like densities are then compared to their analogues in the
continuous\mbox-potential system, and the computational efficiency of
the DMD approach is contrasted with standard molecular dynamics (MD)
simulations. Second, a discontinuous model of a liquid of rigid
molecules of benzene is designed to compare to a united\mbox-atom
model of rigid molecules interacting with continuous potentials.
Equilibrium and dynamical properties of both models are compared under
similar conditions of density and temperature.  The dynamics of
orientational degrees of freedom is examined and the few discrepancies
observed in details of the angular motion are explained.

The paper is organized as follows: The methane system is studied
in Sec.~\ref{Methane}, starting with a discussion of general
considerations pertinent for performing DMD as well as
continuous\mbox-potential simulations of a symmetric\mbox-top, rigid molecular
system in part A, followed by the presentation of gas phase results,
the liquid phase simulation results and analysis of computational
efficiency in parts B, C and D, respectively. In Sec.~\ref{Benzene},
simulations of a united\mbox-atom model of rigid benzene are presented.
Continuous and discontinuous potential models are presented in part A,
followed by a discussion of the simulation details, and the
equilibrium properties results in parts B and C, respectively. A
detailed analysis of the dynamical results, including a careful
examination of the discrepancies associated with the
differences between the models, is provided in part D
of this section.  Finally, conclusions are discussed in
Sec.~\ref{Conclusions}.

\section{Rigid Methane}
\label{Methane}
\subsection{Specification of the model}
\label{methaneModel}

Consider a system of rigid methane molecules in which the four
hydrogen atoms are held fixed with bond length $d$ in a tetrahedral
arrangement around the central carbon atom.  If the mass of the
hydrogen atoms is taken to be the mass of one proton, $m_p$, and the
mass of the central carbon is $12$~$m_p$, the principal moments of
inertia of the rigid rotor system are all equal and given by
$I_1=I_2=I_3=8 m_p d^2/3$.  As discussed in Sec.~II of the preceding
article\cite{preceeding}, statistical averages of a dynamical variable
$X(\bm r_N, \bm p_N)$, depending on the phase space coordinate $(\bm
r_N, \bm p_N)$, for the constrained system are defined by
\begin{eqnarray}
\langle X (\bm r_N,\bm p_N) \rangle 
&=& \frac{1}{\mathcal Z} \int d\bm p_N
d\bm r_N \, \|
\mathbf Z(\bm r_N) \| \nonumber\\
&&\times\,
X(\bm r_N, \bm p_N) \,
\rho(\bm r_N,\bm p_N) \nonumber \\
&&\times \prod_\alpha
\delta( \sigma_\alpha (\bm r_N)) \delta \bigl(
\dot{\sigma}_\alpha(\bm r_N,\bm p_N) \bigr),
\label{statAverage}
\end{eqnarray}
where $\rho(\bm r_N,\bm p_N)$ is the probability density for the
unconstrained system{, $\sigma_\alpha(\bm r_N)=0$ are a set of
$3n-6$ constraint conditions required to hold all distances between
the $n$ atoms in the molecule fixed,} and $\mathcal Z$ is the
partition function for the constrained system, given by
\begin{eqnarray*}
\mathcal Z &=& \int d\bm p_N
d\bm r_N \, \|
\mathbf Z(\bm r_N) \| \, \rho(\bm r_N,\bm p_N)
\nonumber\\
&&\times
 \prod_\alpha
\delta( \sigma_\alpha (\bm r_N)) \delta \left(
\dot{\sigma}_\alpha(\bm r_N,\bm p_N) \right).
\label{partitionFunction}
\end{eqnarray*}
Furthermore, in Eq.~(\ref{statAverage}), the matrix $\mathbf Z(\bm
r_N)$ is given by:
\begin{eqnarray}
\mathbf Z_{\alpha \beta}(\bm r_N) 
&=& \sum_i \frac{1}{m_i} \deriv{\sigma_\alpha
(\bm r_N)}{\bm r_i} \cdot 
\deriv{\sigma_\beta (\bm r_N)}{\bm r_i} .
\label{Zdef}
\end{eqnarray}
For this system, there are $9$ total constraint conditions for
each molecule and they can be chosen to be:
\begin{eqnarray*}
\sigma_1 &=& ( r_{12}^{2} - d^2 )/2 \; \; \; \;  \sigma_6 = (
r_{24}^{2} - a^2 )/2 \\
\sigma_2 &=& ( r_{13}^{2} - d^2 )/2 \; \; \; \; \sigma_7 = (
r_{25}^{2} - a^2 )/2 \\
\sigma_3 &=& ( r_{14}^{2} - d^2 )/2 \; \; \; \; \sigma_8 = (
r_{34}^{2} - a^2 )/2 \\
\sigma_4 &=& ( r_{15}^{2} - d^2 )/2 \; \; \; \;\sigma_9 = (
r_{35}^{2} - a^2 )/2 \\
\sigma_5 &=& ( r_{23}^{2} - a^2 )/2 ,
\end{eqnarray*}
where the carbon atom is numbered $1$, numbers $2$ to $5$ refer to to
hydrogen atoms, and $a = 2d\sqrt{6}/3$ is the distance between
vertices in the tetrahedral structure.  It is evident that the matrix
$\mathbf Z$ of Eq.~(\ref{Zdef}) is a $9 \times 9$ square matrix whose
elements depend on the fixed distances $r_{ij}$ between atoms in the
molecule, as well as angles between them via dot products such as $\bm
r_{12} \cdot \bm r_{13}$.  Since the evolution of the system holds
these values fixed, the $\mathbf Z$ matrix is constant, and is
invertible.  Given that the $\mathbf Z$ matrix is constant, its
determinant acts as a constant multiplicative factor for both the
integral and the normalization $\mathcal Z$ in
Eq.~\eqref{statAverage}, and hence factors out.

For this rigid molecular model of methane, discontinuous interaction
potentials can be constructed by combining repulsive hard\mbox-core and
attractive square\mbox-well interactions as follows. 
The repulsive hard\mbox-core interaction
potential between site $i$ on molecule $a$ and site $j$ on molecule
$b$ is given by
\begin{equation}
V^{\text{rep}}_{ij}\left( r_{ij}^{ab} \right) =
\begin{cases}
\infty & \text{if } r_{ij}^{ab} \leq d_{ij} \\
\ 0     & \text{if } r_{ij}^{ab} > d_{ij},
\end{cases}
\label{rigidmodel}
\end{equation}
where we have used the notation $r_{ij}^{ab}=|\bm r_i^a-\bm r_j^b|$ as
the distance between site $i$ of molecule $a$ and site $j$ of molecule
$b$, and $d_{ij}$ is the interaction distance.  We will use the
natural choice
$d_{23}=d_{24}=d_{25}=d_{34}=d_{35}=d_{45}
\equiv d_{\rm HH}$ (i.e., all hydrogen atoms interact the same
way with one another) and $d_{12} = d_{13} =
d_{14}=d_{15}=d_{\rm CH}$ (i.e., all hydrogen atoms
interact the same way with carbon atoms), and we will also denote
$d_{11}=d_{\rm CC}$.  Choosing also $d_{\rm CH} =
\frac{1}{2}(d_{\rm CC} + d_{\rm HH})$, the values of
$d_{\rm CC}$ and $d_{\rm HH}$ can be interpreted as the
diameter of the carbon and hydrogen atom, respectively. However, a
more general case is possible where the atom's effective diameter is
dependent on the identity of its interaction partner.

In addition to the hard\mbox-core repulsive interaction potentials above,
we include an attractive square well ($SW$) between carbon atoms in
the model of the form
\begin{equation}
V_{11}^{\rm attr}\left( r_{11}^{ab}  \right) =
\begin{cases}
-V_{SW}        & \text{if } r_{11}^{ab} \leq d_{SW} \\
\ \ \ 0        & \text{if } r_{11}^{ab} > d_{SW},
\end{cases}
\label{squareWell}
\end{equation}
where the values of $V_{SW}$ and $d_{SW}$ are adjustable
parameters and the obvious condition $d_{SW} > d_{\rm CC}$
must hold for this potential to have any effect.  The attractive
potential allows for a condensation transition in a dense fluid of
methane molecules at low temperature.

The DMD results for this methane model presented below were obtained
with the following parameter values: $d_{\rm CC}=3.3$~\AA, $d_{\rm
HH}=2.75$~\AA, $d_{\rm CH}=3.025$~\AA, $d_{\rm SW}= 5.1$~\AA\ and
$V_{SW}=1.824$~kJ/mol.

Given the form of the potentials in the model, the {\it exact}
trajectory can be computed from arbitrary initial
conditions.\footnote{This exactness applies to the free flight
portions of the motion and the collision rules. The time of a
collision event cannot be expressed in an exact closed form in
general, yet it can be determined to arbitrary precision (in practice
machine precision) using any of a number of numerical
procedures\cite{NumRecipes}}\nocite{NumRecipes} At any time $t$, the
orientation of a given methane molecule is specified by a rotation
matrix $\mathbf A(t)$, known as the {\it attitude matrix}, while its
center of mass undergoes free linear motion in between impulsive
events that alter momenta in a discontinuous fashion.  The attitude
matrix and its transpose $\mathbf A^\dagger (t)$ can be used to
transform between the lab and principal axis (body) frames and to
calculate the Cartesian positions $\bm r_i(t)$ of an atom $i$ in a
molecule using
\begin{eqnarray}
\bm r_i(t) 
&=& \bm R ( t)+\mathbf A^\dagger(t) \cdot \tilde{\bm r}_i \nonumber \\ 
&=& \bm R(0) + \bm V(0) t +\mathbf A^\dagger(t) \cdot \tilde{\bm r}_i,
\label{relcartesianpositions}
\end{eqnarray}
where $\tilde{\bm r}_i$ is the constant position vector in the body
frame, and $\bm R$ and $\bm V$ are the Cartesian vectors in the
laboratory frame for the center of mass position and velocity,
respectively.
 
As discussed in detail in Ref.~\onlinecite{preceeding}, the time
dependence of the attitude matrix is given by
\begin{equation}
 \mathbf  A(t) = \mathbf P(t)\cdot \mathbf A(0).
\label{Asolution}
\end{equation}
where $\mathbf P(t)$ is a rotation matrix itself which `propagates'
the orientation $\mathbf A(0)$ to the orientation at time $t$.  For a
spherical rotor in which all the principal moments of inertia are
equal, the propagator matrix $\mathbf P(t)$ is particularly simple,
\begin{align}
  \mathbf P(t) &=
  \mathbf U(-\tilde{\bm \omega} t),
                                 \label{Pmatrix}
\end{align}
where $\tilde{\bm \omega}$ is the angular velocity vector in the body
frame.  In Eq.~(\ref{Pmatrix}), $\mathbf U
(-\tilde{\bm\omega}t)=\mathbf U(\theta\hat{\bm n})$ is a rotation
matrix rotating an arbitrary vector by an angle $\theta=-|\bm \omega| t$
around an axis $\hat{\bm n}=\bm \omega/|\bm \omega|$ with components
$(n_1,n_2,n_3)$ according to
\begin{widetext}
\begin{equation}
\mathbf U(\theta\hat{\bm n}) 
=
\begin{pmatrix}
n_1^{2} + \left( n_2^2+n_3^2 \right) \cos\theta &
n_1 n_2 \left( 1 - \cos\theta \right) - n_3\sin\theta &
n_1 n_3 \left( 1 - \cos\theta \right) + n_2\sin\theta \\
 n_1 n_2 \left( 1 - \cos\theta \right) + n_3\sin\theta &
n_2^{2} + \left( n_1^2+n_3^2 \right) \cos\theta &
 n_2 n_3 \left( 1 - \cos\theta \right) - n_1 \sin\theta \\
 n_3 n_1 \left( 1 - \cos\theta \right) - n_2 \sin\theta &
 n_3 n_2 \left( 1 - \cos\theta \right) + n_1 \sin\theta &
n_3^{2} + \left( n_1^2+n_2^2 \right) \cos\theta 
\end{pmatrix}.
\label{generalRotation}
\end{equation}
\end{widetext}
For a spherical rotor, it follows from the Euler equations that
$\tilde{\bm \omega}$ is a constant vector determined by initial
conditions, and hence the matrix $\bm U (-\tilde{\bm \omega} t)$ can be
viewed as a rotation around a fixed axis by an angle that is a linear
function of time.  Note that the moment of inertia tensor in the
laboratory frame $\mathbf I (t)$ and in the principal axis frame
$\tilde{\mathbf I}$ are related by $\mathbf I(t) = \mathbf A^\dagger
(t) \cdot \tilde{\mathbf I} \cdot \mathbf A(t)$, with
$\tilde{\mathbf I} =\text{diag}(I_1,I_1,I_1)$ for a spherical rotor,
implying that $\mathbf I(t)$ is diagonal,
constant, and equal to the moment of inertia tensor in the principal
axis frame $\tilde{\mathbf I}$.

The angular momentum vector $\tilde{\bm \omega}$, defining the axis
of rotation in $\bm P(t)$, remains constant until an
impuls{ive force acts on it}\ during a collision event.  The times
$t_c$ at which the impulses act are determined by zeros of a collision
indicator function $f_{ij}(t_c) = r_{ij}^{ab} - d_{ij}$,
describing the time at which the distance $r_{ij}^{ab}$ between atom
$i$ of body $a$ and atom $j$ of body $b$ attains a value at which
there is a discontinuity in the potential energy (see
Eq.~(\ref{rigidmodel})).  As explained in
Ref.~\onlinecite{preceeding}, the collision times can be computed
using numerical grid\mbox-search methods using
Eq.~(\ref{relcartesianpositions}).

As discussed in Ref.~\onlinecite{preceeding}, the effect of the
impulses on the Cartesian coordinates or angular velocities can be
computed either from a constrained variable approach, using the
$\bm Z$ matrix, or using a rigid\mbox-body approach. In the
rigid\mbox-body approach the impulse at collision time $t_c$ leads to a
discontinuity in the center of mass momentum vector $\bm P_a$ and
angular velocity vector $\bm \omega_a$ of body $a$ in the laboratory frame
according to
\begin{eqnarray}
\bm P'_a &=& \bm P_a + \Delta \bm P_a 
\label{postPCOM}\\
\bm \omega_a' &=& \bm \omega_a + \Delta \bm \omega_a,
\label{postOmega}
\end{eqnarray}
where $\bm P_a = M\bm V_a$ is the center of mass momentum vector of
body $a$, $M$ is the total mass of body $a$, and $\bm P_a'$ and
$\bm \omega_a'$ indicate the post\mbox-collisional center of mass
momentum and angular velocity vectors.  Application of conservation
laws leads to the following results:
\begin{eqnarray}
\Delta \bm P_a &=&   -S\hat{\bm r}_{ij}^{ab}  \label{changePCOM}\\
\Delta \bm \omega_a &=& -S\, \mathbf I_a^{-1} \cdot \left(
\overline{\bm r}^a_i \times \hat{\bm r}^{ab}_{ij} \right) ,
\label{changeOmega}  
\end{eqnarray}
where $S$ is the magnitude of the impulse given by
\begin{equation}
S = \frac{-b \pm \sqrt{b^2 - 4a\Delta \Phi}}{2a},
\label{magImpulse}
\end{equation}
$\Delta \Phi$ is the height of the discontinuity in the potential at the
interaction distance,
\begin{eqnarray}
a &=& \frac{1}{2M_a} + \frac{1}{2M_b} 
+ \frac{\Delta E^a_\omega + \Delta
E^{b}_\omega}{2} \nonumber \\
b 
&=&  \bm v^{ab}_{ij} \cdot \hat{\bm r}^{ab}_{ij},\nonumber
\end{eqnarray}
and
\begin{eqnarray}
\Delta E^a_\omega &=& \bm n_{ij}^{a\dagger} \cdot \mathbf I_a^{-1} \cdot \bm n^a_{ij}\nonumber \\
\Delta E^{b}_\omega &=& \bm n^{b\dagger}_{ij} \cdot \mathbf I_b^{-1} \cdot \bm n^{b}_{ij},
\nonumber
\end{eqnarray}
with 
\begin{eqnarray*}
  \bm n^a_{ij} &=& \overline{\bm r}_i^a \times \hat{\bm r}_{ij}^{ab}
\\
\bm n^b_{ij} &=& \overline{\bm r}_j^{b} \times \hat{\bm r}_{ij}^{ab}.
\end{eqnarray*}
In the equations above, all quantities are to be taken at time
$t=t_c$ and $\overline{\bm r}_i^a = \bm r_i^a - \bm R^a$ denotes the
position vector of atom $i$ on molecule $a$ relative to its center of
mass $\bm R^a$ and $\hat{\bm r}_{ij}^{ab}$ is the unit vector along
the direction of the vector $\bm r_j^b - \bm r_i^a$ connecting atom
$i$ on body $a$ with its colliding partner $j$ on body $b$.  The
physical solution of $S$ corresponds to
the positive (negative) root if $b >0$ ($b<0$), provided $b^2 >
4a\Delta \Phi$.  If this latter condition is not met, there is not
enough kinetic energy to overcome the discontinuous barrier and the
system experiences a hard\mbox-core scattering, in which case
$\Delta\Phi$ is replaced by zero and consequently $S=-b/a$. For the
spherical rotor methane system, the fact that the moment of inertia
tensor in the lab frame is proportional to the identity matrix leads
to the simplification $\Delta E^{a,b}_\omega = \bm n_{ij}^{a,b
\dagger} \cdot \bm n^{a,b}_{ij}/I_1$.

With this model in hand, a DMD simulation can be carried out where
four different types of events can take place, namely: hard\mbox-core
collisions; square\mbox-well interactions; cell crossings; and virtual
collisions associated with the truncation of numerical searches for
collision events (described in detail in
Ref.~\onlinecite{preceeding}).  While the time at which cell crossings
and carbon\mbox-carbon repulsive and attractive interaction events occur
can be calculated analytically from the linearity of the
center\mbox-of\mbox-mass motion, the times for the other atom\mbox-atom
intermolecular interactions must be computed numerically using the
grid\mbox-search method elaborated in Ref.~\onlinecite{preceeding}.  The
use of cell subdivisions, local clocks, and a binary tree to manage
the event calendar is a standard practice in this type of simulation
and largely improves the simulation's efficiency\cite{Rapaport}.

For the purpose of comparison, a rigid molecular model of methane
based on continuous interaction potentials was also implemented. In
this model, all intermolecular interactions were assumed to be of
Lennard\mbox-Jones form:
\begin{equation}
V(r_{ij}) =  \epsilon_{ij} \left[ \left( \frac{R_{\rm min}^{ij}}{r_{ij}} \right)^{12} - 2 
\left( \frac{R_{\rm min}^{ij}}{r_{ij}} \right)^{6} \right] ,
\label{softmodel}
\end{equation}
where $R_{\rm min}^{ij} = \frac{1}{2}(R_{\rm min}^{i}+R_{\rm
min}^{j})$ and $\epsilon_{ij}=\sqrt{\epsilon_i\epsilon_j}$. Here,
$\epsilon_{i}$ and $R_{\rm min}^i$ represent respectively the value
and the position of the minimum of the Lennard\mbox-Jones potential
between atoms of the same kind, i.e. $V(r_{ii})$.  The parameters
$\epsilon_{i}$ and $R_{\rm min}^{i}$ are taken from
Ref.~\onlinecite{charmm}, where these and other parameters have been
computed for atoms in a wide variety of molecules of biological
interest and are intended to be used in standard condensed\mbox-phase
simulations.  The values of $d_{ij}$ in the DMD model were in fact
chosen to be comparable to $R_{\rm min}^{i}$.  Standard values were
used for all parameters in the continuous potential system: the
C\mbox-H bond distance is $1.11$ \AA, $R_{\rm min}^{\rm CC}=3.60$~\AA,
$R_{\rm min}^{\rm CH}=3.13$~\AA, $R_{\rm min}^{\rm HH}=2.66$~\AA,
$\epsilon_{\rm CC}=0.388$~kJ/mol, $\epsilon_{\rm CH}=0.260$~kJ/mol,
and $\epsilon_{\rm HH}=0.176$~kJ/mol.

The continuous MD simulations were carried out in a standard
fashion. The equations of motions were integrated with a 4th order
predictor\mbox-corrector algorithm with rotations represented by
quaternion parameters. For efficiency, the forces and torques were
calculated using a cell structure where the index of molecules in each
cell were stored in a linked list.  In the simulation, the cell size
was set to be equal to the cut\mbox-off value of the
Lennard\mbox-Jones potential, taken to be 6.18 \AA.

\subsection{Low density results}
\label{lowdensity}

\begin{figure}[b]
\includegraphics[width=0.85\columnwidth,angle=0,clip]{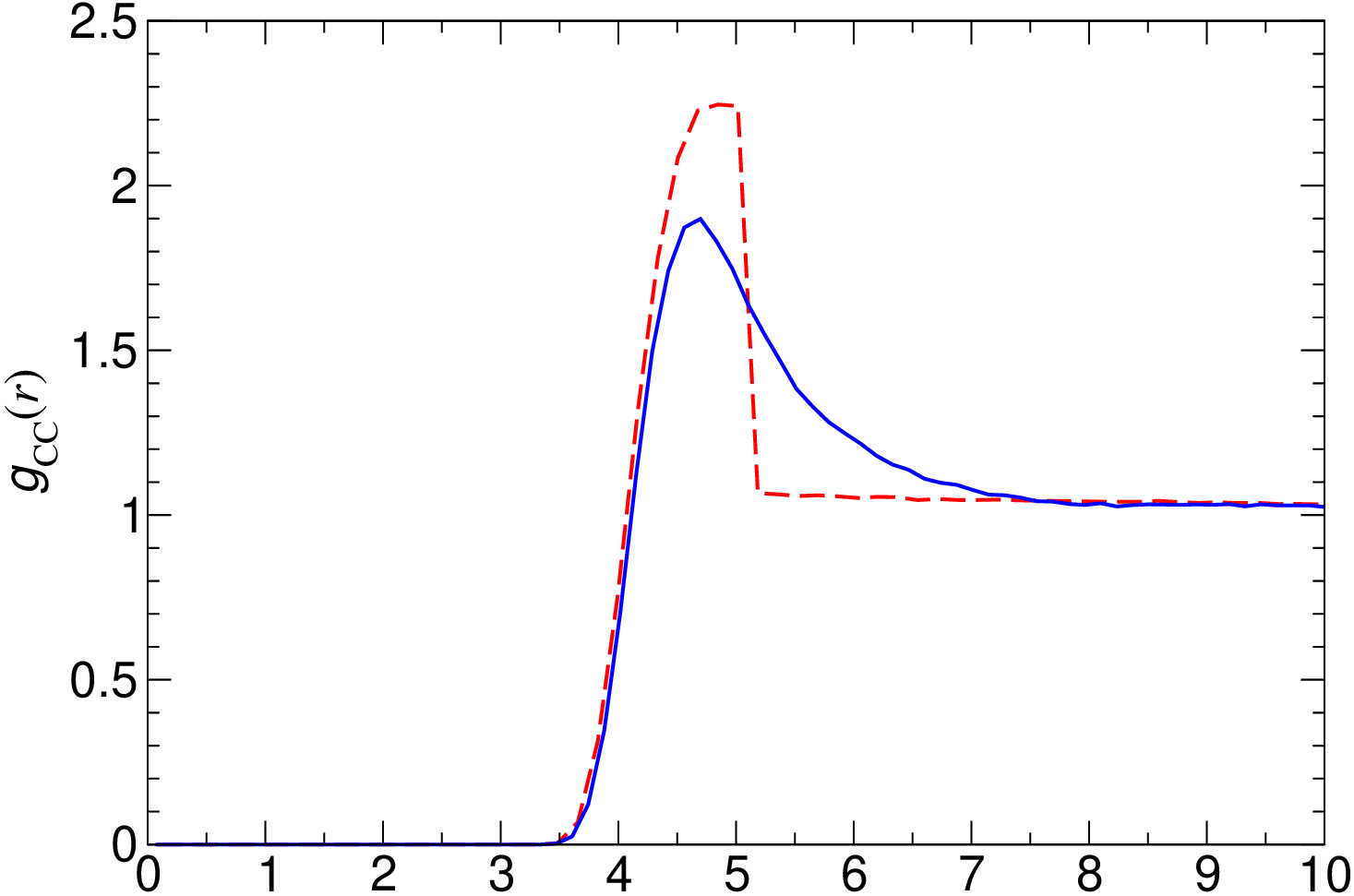}
\includegraphics[width=0.85\columnwidth,angle=0,clip]{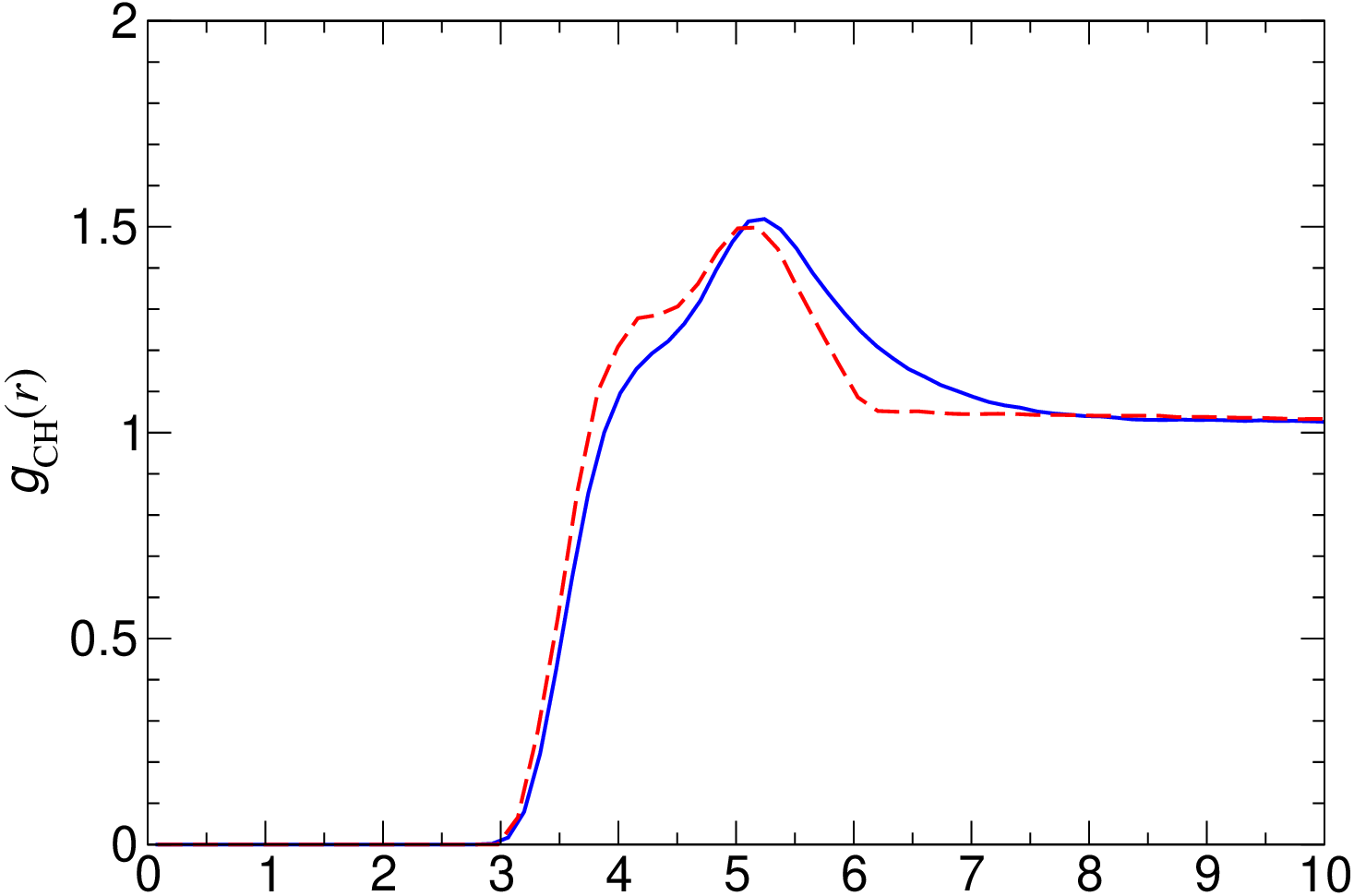}
\includegraphics[width=0.85\columnwidth,angle=0,clip]{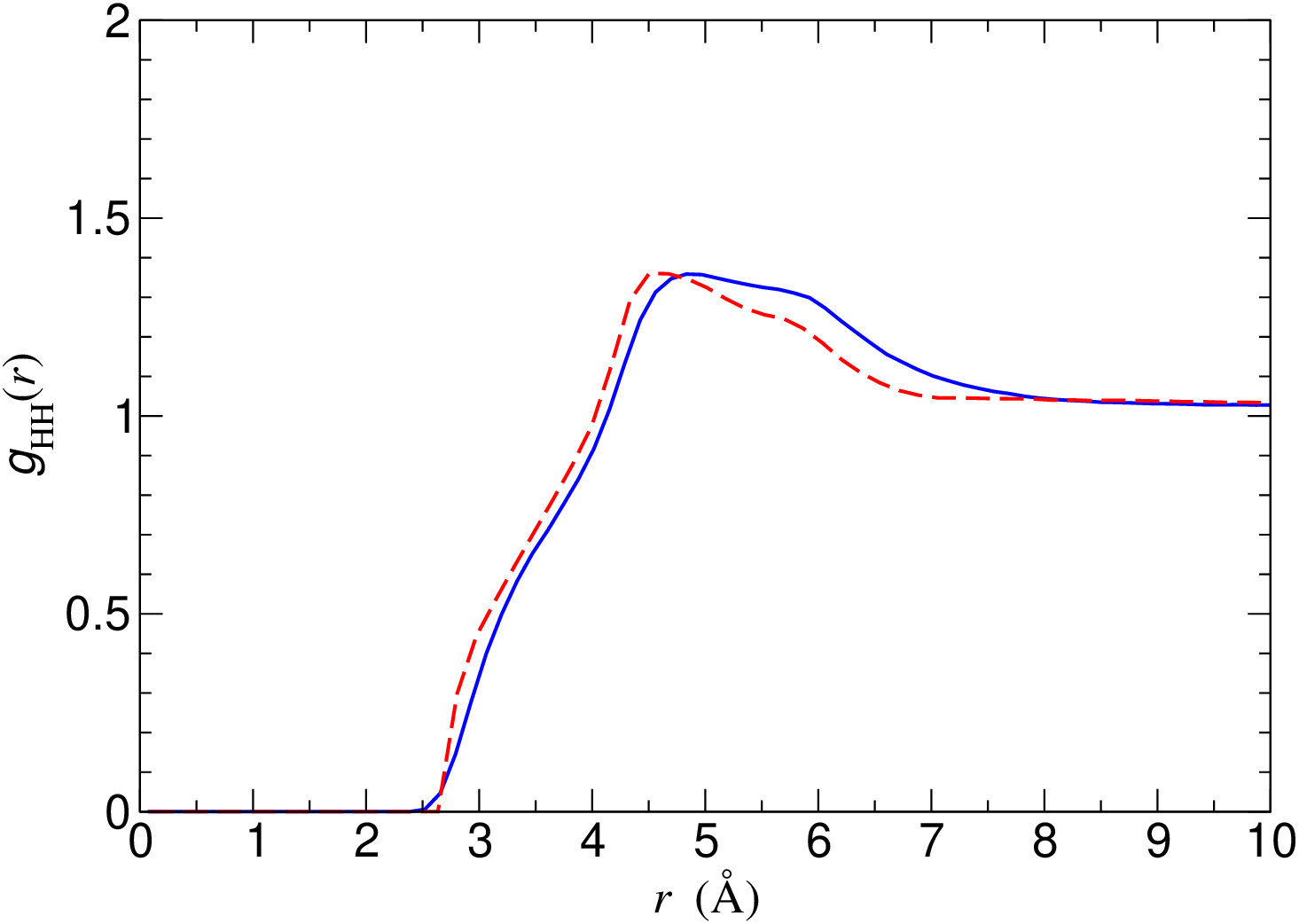}
\caption{From top to bottom, carbon\mbox-carbon, carbon\mbox-hydrogen
and hydrogen\mbox-hydrogen RDFs for methane at a gaseous density of
$3.47\cdot10^{-3}\rm\ g/cm^3$ and a temperature of 298~K.  The
continuous lines represent the continuous MD model and the dashed
lines represent the DMD results.  The parameters of the model are
given in the text.}
\label{fig:radial-ld}
\end{figure}

\begin{table}
\begin{center}
\begin{tabular}{|c|r@{\mbox-}l|c|c|c|c|}
\hline
\ \ \ $AB$ \ \ \ & \multicolumn{2}{c|}{\ $r_1$\mbox-$r_2$ range }& \ \ \ ${q}_{AB}$ (DMD) \ \ & \ \ \ ${q}_{AB}$ (MD) \ \ \  \\ \hline
\hline
CC   &\ \ \ \ 0 & 6.7 \AA          & 0.18 $\pm0.01$ & 0.18 $\pm0.01$  \\ \hline
CH   & 0        & 6.7 \AA          & 0.71 $\pm0.01$ & 0.74 $\pm0.01$  \\ \hline
HH   & 0        & 6.7 \AA          & 0.71 $\pm0.01$ & 0.73 $\pm0.01$  \\
\hline
\end{tabular}
\caption{Number of nearby atoms associated with the various peaks of the
RDFs for the low\mbox-density methane fluid presented in
Fig.~\ref{fig:radial-ld}. The values are calculated using
Eq.~(\ref{neighbors}).}
\label{tbl:lowdens}
\end{center}
\end{table}
  
To study a low density methane system, DMD and
continuous\mbox-potential simulations were performed for the
models introduced above.  In both cases, a system with 512 molecules
was simulated at a temperature $T=298$~K ($k_{\rm B}T =
2.478$~kJ/mol) and a density of $3.47 \cdot 10^{-3} \rm\ g/cm^3$.

Temperature is defined here as twice the kinetic
energy divided by the number of degrees of freedom in the system. In
both types of simulation, initial points in phase space were drawn
from a so-called iso\mbox-kinetic ensemble where the kinetic energy is
fixed according the temperature, while the potential energy can vary.
In the DMD simulations, however, it turns out that the resulting
fluctuations in the potential energy are so small (less than half of a
percent) that a trajectory at one fixed energy may be used as
representative of the whole iso-kinetic ensemble.

In order to give correct time\mbox-correlation functions, the
dynamics calculated in the simulations should in principle take place
at constant total energy. However, the continuous\mbox-potential
simulation exhibits a systematic energy drift, which would violate the
conservation of energy. Therefore, it was supplied with a thermostat
in the form of a periodic and infrequent (e.g.\ every picosecond)
rescaling of the velocities. For consistency a similar infrequent
rescaling of the velocities could be performed in the DMD simulations,
but this has a negligible effect because of the above mentioned small
fluctuations of the potential energy in the DMD simulation.

The DMD simulation, after equilibration, was run at constant energy of
$7.37 \rm\ kJ/mol$. The pressure was calculated using the standard
impulsive limit formula\cite{Rapaport} and was found to be $5.303 \pm
0.009 \rm\ bar$.  The continuous MD simulation was performed at
constant temperature by using the above mentioned
velocity\mbox-rescaling thermostat.  The total energy and pressure
values were $7.236 \pm 0.05 \rm\ kJ/mol$ and $10.44 \pm 0.09 \rm\
bar$, respectively.

Although ideally, one would like to simulate a system at standard
pressure (1~bar), the pressure in a simulation at constant volume and
energy is very sensitive to small changes in the parameters of the
interaction potential. This is true for both the DMD and the standard
continuous potential simulation technique. For this reason, pressures
which do not differ by an order of magnitude (such as those above) may
be considered as being in reasonable agreement.

In Fig.~\ref{fig:radial-ld}, the intermolecular carbon\mbox-carbon,
carbon\mbox-hydrogen, and hydrogen\mbox-hydrogen radial distribution functions
(RDFs) obtained from the DMD simulation are shown as a function of
radial distance, and compared to those obtained from the continuous
simulation.  From the lack of correlation evident in these RDFS, it is
clear that the structure of the fluid is that of a gas, which is the
expected phase at this density and temperature. Note that the
agreement between the carbon\mbox-hydrogen and hydrogen\mbox-hydrogen RDFs is
quite good, while the carbon\mbox-carbon RDF exhibit a clear difference in
the shape of the only peak that appears. The difference between the
two models in the peak structure for the carbon\mbox-carbon RDF is to be
expected since there is no interaction between molecules separated by
a distance greater than the square\mbox-well parameter (5.1 \AA) in the DMD
model.  In the continuous model, on the other hand, the interaction
distance extends up to 6.18 \AA.

A quantitative comparison of the RDFs can be carried out by
calculating the average number $q_{AB}$ of atoms of type $B$ that are
close neighbors of an atom of type $A$, where $q_{AB}$ is given by
\begin{equation}
q_{AB}  = 4 \pi \rho_B \int_{r_1}^{r_2} r^2 g_{AB}(r) dr, 
\label{neighbors}
\end{equation}
where $\rho_B$ is the number density of type $B$ and $g_{AB}(r)$ is
the radial distribution function for the pair $AB$.  With the choice
of $r_1=0$ and $r_2 = 6.7$ \AA, one finds that $q_{CC} = 0.18$ from
the carbon\mbox-carbon RDF for both the DMD and continuous\mbox-potential
models, indicating that the average number of near neighbor molecules
around a given molecule is essentially the same in the two models at
low density.  A similar quality of agreement of all $q_{AB}$ is
observed for all other pair types, as is evident in
Table~\ref{tbl:lowdens}.

\subsection{High density results}
\label{highdensity}

\begin{figure}[b]
\includegraphics[width=0.85\columnwidth,angle=0,clip]{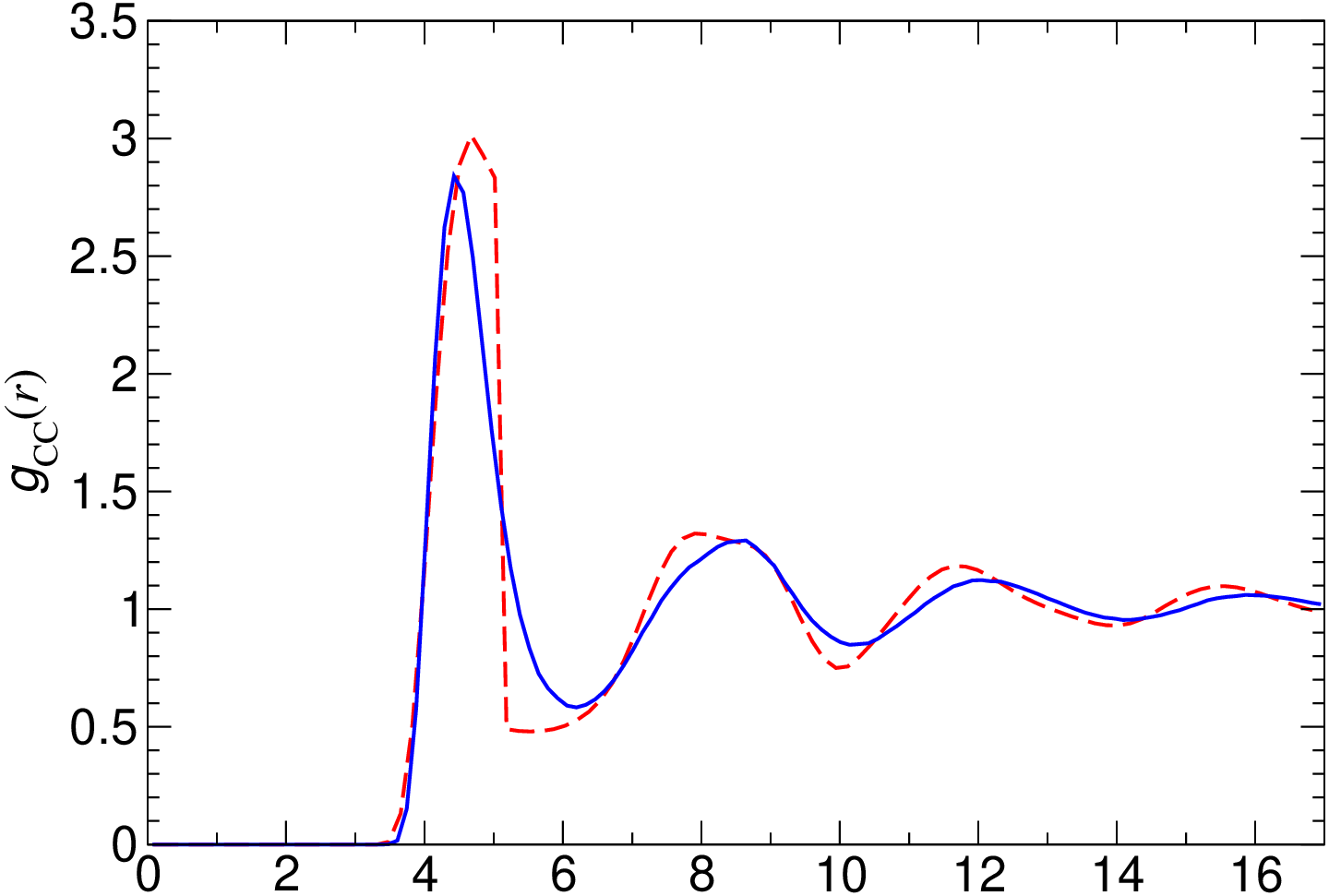}
\includegraphics[width=0.85\columnwidth,angle=0,clip]{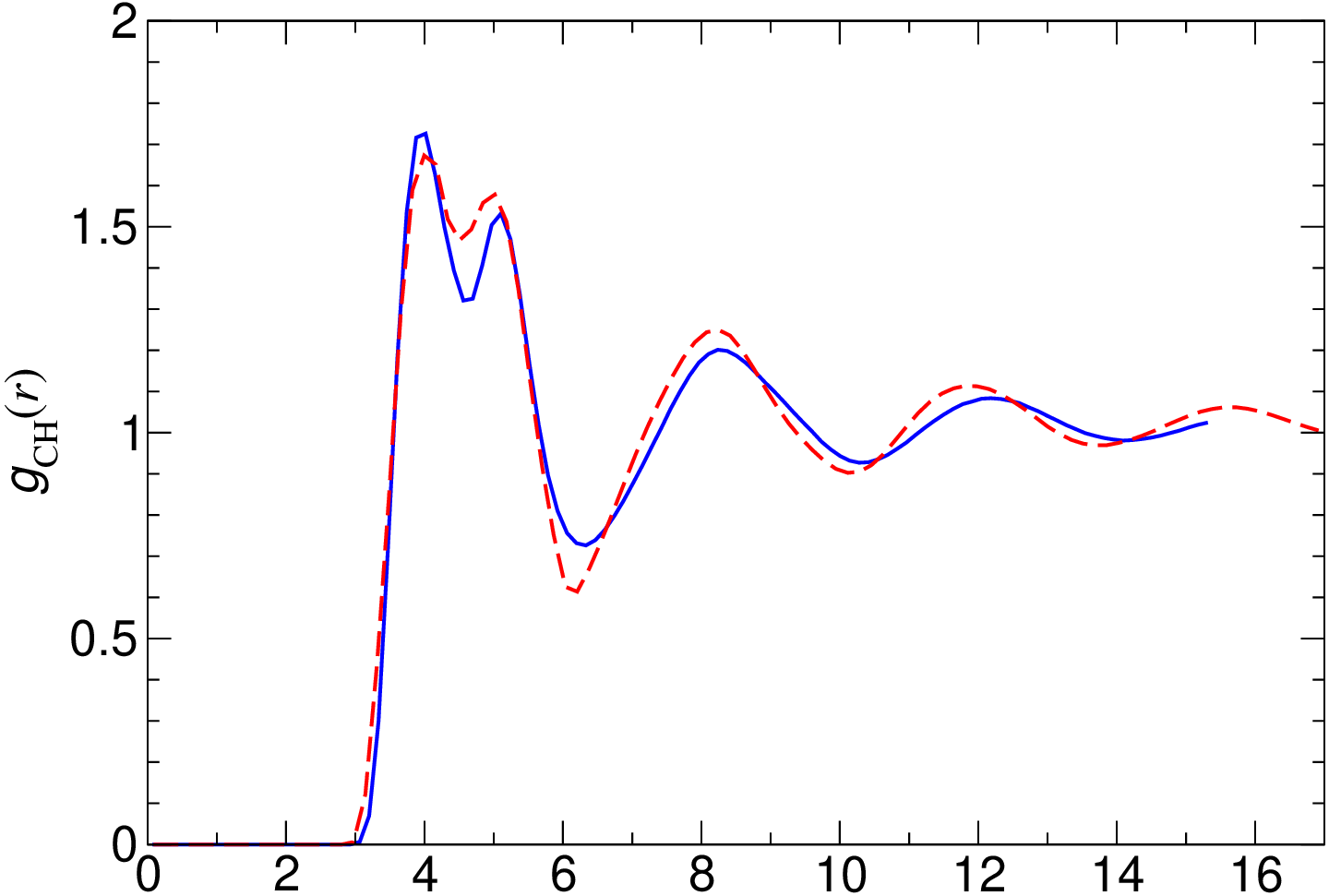}
\includegraphics[width=0.85\columnwidth,angle=0,clip]{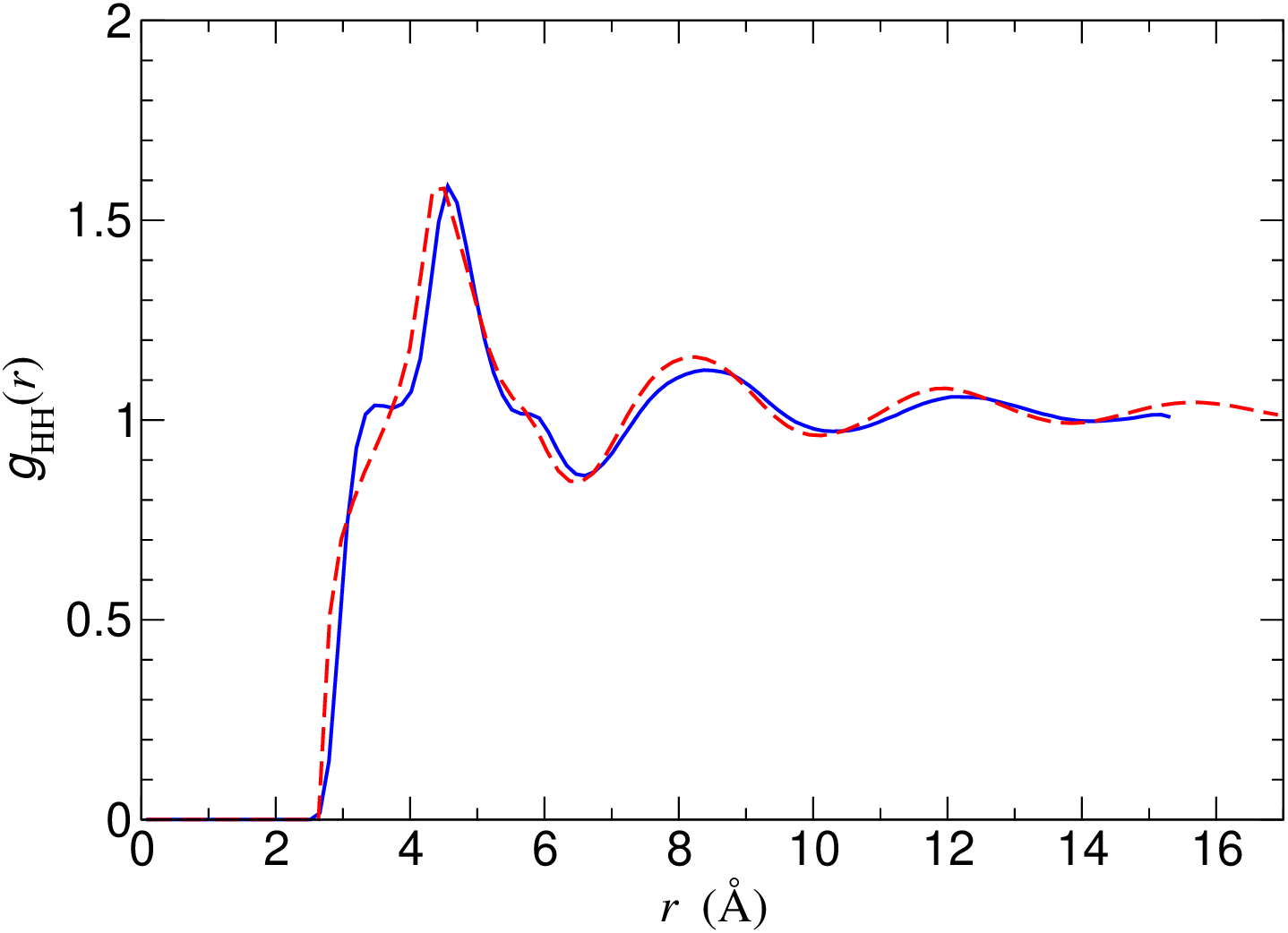}
\caption{From top to bottom, carbon\mbox-carbon, carbon\mbox-hydrogen, and
hydrogen\mbox-hydrogen RDFs for methane at a liquid\mbox-like density of
$0.347\rm\ g/cm^3$ and a temperature of 126.8~K. The
continuous lines represent the continuous MD model and the dashed
lines represent the DMD results.  The model's parameters are given
in the text.}
\label{fig:radial-hd}
\end{figure}

A more challenging test for the discontinuous model is in the high
density and moderately low temperature regime where the system
exhibits liquid behaviour and more complex structure. The conditions
simulated here correspond to a density of $0.347 \rm\ g/cm^3$ and a
temperature of $126.8\rm\,K$ ($k_{\rm B}T = 1.064$~kJ/mol).  Under
these conditions, the average total energy in the DMD and
continuous\mbox-potential model simulations was found to be $-5.583
\rm\,kJ/mol$ and $-6.11 \pm 0.05 \rm\,kJ/mol$, respectively, with
corresponding pressures of $61 \pm 10 \rm\,bar$ and $102 \pm 58
\rm\,bar$ (which are in reasonable agreement in the sense explained
above).

In Fig.~\ref{fig:radial-hd}, the intermolecular carbon\mbox-carbon,
carbon\mbox-hydrogen and hydrogen\mbox-hydrogen RDFs are plotted versus radial
distance.  The correlation functions show a highly\mbox-ordered fluid
and are consistent with a liquid phase.  Note that most details in the
RDFs obtained for the continuous model are well reproduced in the DMD
model.

For a quantitative comparison, we again examine the average number of
near neighbors $q_{AB}$, as defined in
Eq.~(\ref{neighbors}). The results are presented in
Table~\ref{tbl:angles}. From the values associated with the first peak
in the carbon\mbox-carbon RDFs, we see that every molecule is surrounded
on average by about $13$ other methane molecules in both
models. Furthermore, the second peak integrates to about $46$, a value
that is slightly larger than the value of $44$ found with the
continuous model. However, given that this distance is well beyond the
square\mbox-well interaction range, the agreement is quite good and
suggests that excluded volume effects are the predominant factor in the
structural packing in this simple liquid system, while details of
specific interactions are relatively unimportant.

It also is worth noting that the agreement between the models found in
the structural functions at both densities for the same temperature
and similar pressure values suggests that they would have a similar
phase diagram. An exhaustive analysis of the phase diagram of this
model is, however, beyond the scope of this paper.

\begin{table}
\begin{center}
\begin{tabular}{|c|r@{\mbox-}l|c|c|c|c|}
\hline
\ \ \ $AB$ \ \ \ & \multicolumn{2}{c|}{\ $r_1$\mbox-$r_2$ range }& \ \ \ $q_{AB}$ (DMD) \ \ & \ \ \ $q_{AB}$ (MD) \ \ \  \\ \hline
\hline
CC   & 0 & 6.19 \AA            & 12.9$\pm$0.05    & 13.4$\pm$0.05   \\ \hline
CC   &\  6.19 & 10.14 \AA \    & 45.7$\pm$0.05    & 43.6$\pm$0.05   \\ \hline
CH   & 0 & 6.33 \AA            & 56.7$\pm$0.05    & 53.7$\pm$0.05   \\ \hline
HH   & 0 & 6.6  \AA            & 58.0$\pm$0.05    & 59.6$\pm$0.05   \\
\hline
\end{tabular}
\caption{Number of nearby atoms associated to the various
peaks of the RDFs for the methane fluid presented in
Fig.~\ref{fig:radial-hd}. The values are calculated using
Eq.~(\ref{neighbors}).}
\label{tbl:angles}
\end{center}
\end{table}

\begin{figure}[b]
\centerline{\includegraphics[angle=0,width=0.85\columnwidth]{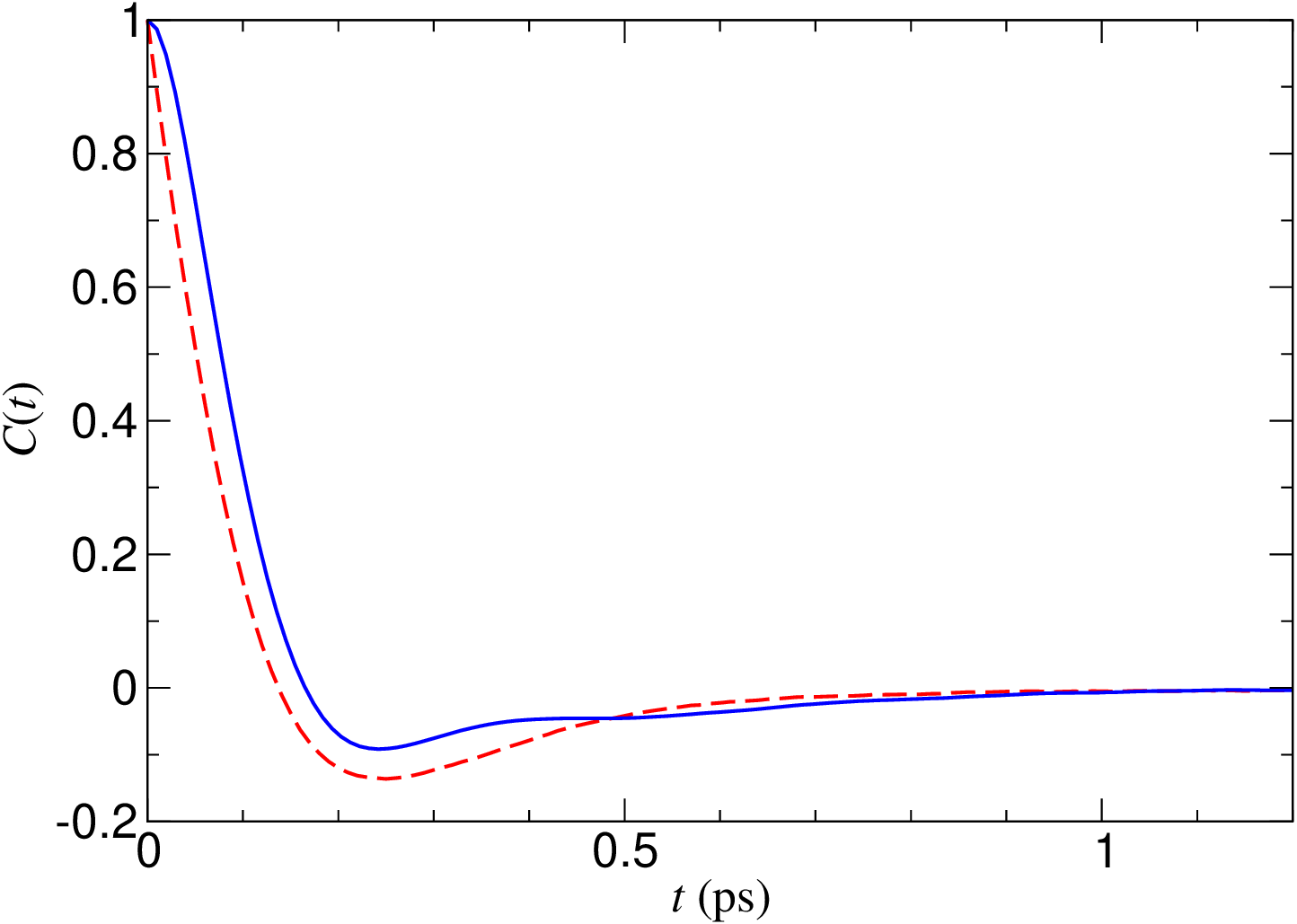}}
\caption{Velocity time correlation function for the case of
liquid methane of Sec.~\ref{highdensity}.  The continuous and
dashed lines correspond to the MD and DMD results, respectively.}
\label{velAcf}
\end{figure}

In addition to the structural properties of the DMD model,
event\mbox-driven simulations also enable dynamical correlations to be
computed in rigid\mbox-body systems.  In fact, unlike continuous\mbox-potential
systems, the DMD approach allows dynamical properties to be calculated
from essentially exact (i.e., up to machine precision) trajectories
due to the simplicity of the form of the interaction potential.
Fig.~\ref{velAcf} shows the results obtained for the normalized time
autocorrelation function of the center of mass velocity (VACF) of a
given molecule in the fluid, $C(t)= \langle\bm V(0)\cdot\bm
V(t)\rangle/\langle\bm V(0)\cdot\bm V(0)\rangle$, for the continuous
and discontinuous potential models under liquid conditions.
Considering the fundamental differences in the nature of motion in the
two models, the agreement at a qualitative level of $C(t)$ between the
two models is somewhat surprising.  Nonetheless, it is evident that
the short\mbox-time behavior, in which the VACF in the DMD model decays
more quickly than in the continuous\mbox-potential model, and the degree of
anti\mbox-correlation in the velocities or ``cage'' effect at longer time
scales differ appreciably. 

Because of the qualitative similarity in the VACFs for the two models,
the value of the self\mbox-diffusion coefficient, $D =
(k_BT/m)\int_0^\infty C(t)dt$, agrees reasonably well: $D$ is found to
be $(0.89\pm0.03)\cdot10^{-5}\rm\,cm^2/s$ and
$(3.10\pm0.06)\cdot10^{-5}\rm\,cm^2/s$ for the DMD and continuous MD
model, respectively. The somewhat lower value for the diffusion
constant in the DMD code might be due to the strong
center\mbox-of\mbox-mass attraction in the DMD model, in contrast to
the continuous model in which the attractive potential is distributed
over the atoms.  The stronger attractive potential in the DMD model
may cause the molecules to stay bonded for a longer time, thus
decreasing the mobility.

\subsection{Efficiency}
\label{Efficiency}

\begin{figure}[t]
\centerline{\includegraphics[angle=0,width=0.85\columnwidth]{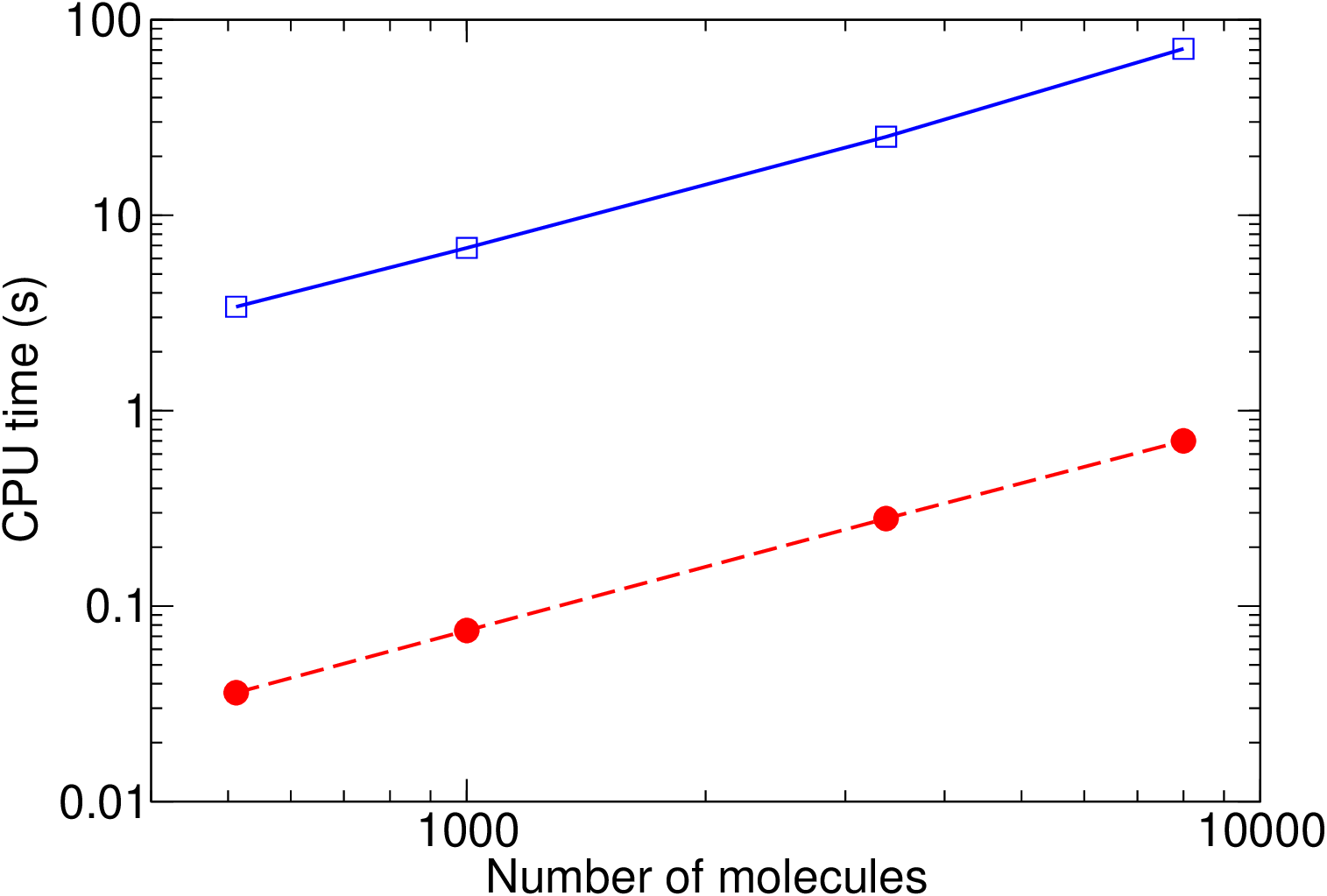}}
\caption{CPU time per picosecond of real time as a function of the
system size for the low density methane case of
Sec.~\ref{lowdensity}.  The dots correspond to the DMD values whereas
the open squares are the values obtained in a continuous MD
simulation. The solid and dashed lines join the DMD and MD values and
show the expected linear scaling with system size.}
\label{efficiency-ld}
\end{figure}

Given the complexity of constructing an event\mbox-driven simulation for
rigid systems, it is natural to wonder whether it is worthwhile to
carry out such calculations. Such concerns can only be addressed by
considering the relative efficiency of event\mbox-driven versus standard
simulations as a function of the system size and of the physical
conditions of the simulation.  In order to assess the relative
computational efficiency of the DMD simulation, the CPU time needed to
simulate one picosecond of real time dynamics was assessed for the DMD
and continuous\mbox-potential models for different system sizes and
densities.  Of course, such a comparison depends sensitively on a
number of factors that have little to do with the methods themselves,
such as the choice of programming language, computer architecture,
compiler choice and degree of compiler optimization.  In order to
minimize such external effects, the DMD code was written in a high
level language (C++) while the code for the continuous\mbox-potential
simulations was adapted from Ref.~\onlinecite{Rapaport} and was
written in C.  Both codes were compiled with the same compiler, on
the same computer cluster, and with the same non\mbox-specialized level of
optimization. Since object\mbox-oriented code in C++ is generally
considered to be less efficient than C because of the additional
overhead of utilizing classes, it is likely that optimizations in
coding style and changing the coding language to C would enhance the
relative efficiency of the DMD code reported here.

\begin{figure}[t]
\centerline{\includegraphics[angle=0,width=0.85\columnwidth]{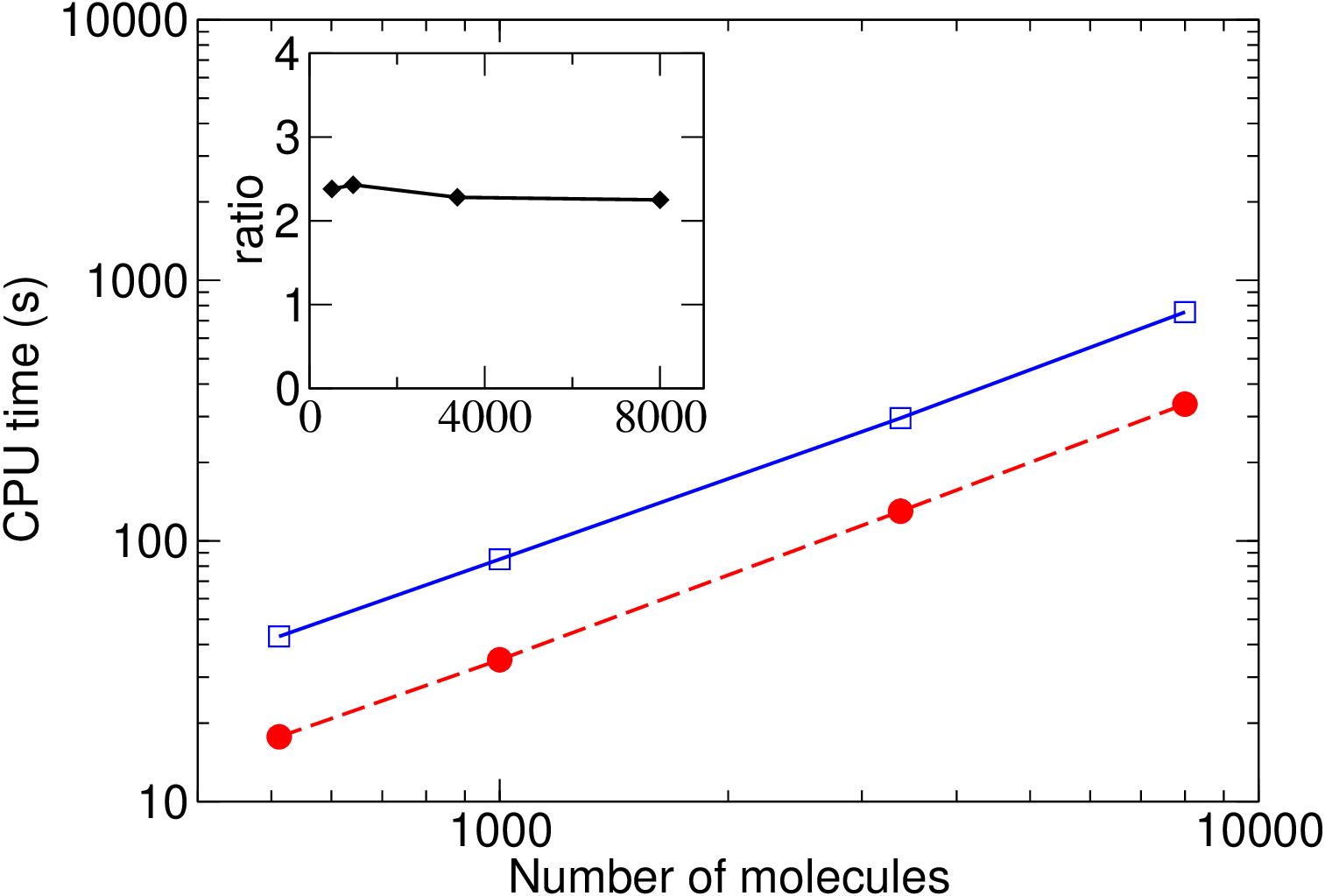}}
\caption{CPU time per picosecond of real time as a function of the
system size for the high density methane case of
Sec.~\ref{highdensity}.  The dots correspond to the DMD results and
the open squares correspond to the continuous MD simulation.  The
insert shows the ratio MD/DMD of the CPU time values as a function of
system size.}
\label{efficiency-hd}
\end{figure}

It is also clear that the efficiency of a continuous MD simulation is
largely dependent on the size of the time step chosen to integrate the
equations of motion, i.e., the larger the time step the more rapidly
the simulation propagates the system.  However, a large integration
time step leads to numerical instabilities in the MD trajectory,
something that can only be tolerated with an extensive use of
thermostats.  The use of thermostated dynamics has its own limitations
since it artificially biases the ``true'' dynamics and influences the
calculation of dynamical properties. The effect on the dynamics of
thermostats is generally unknown, and hence artificial means of
stabilizing trajectories must be handled with great care when looking
at time dependent phenomena. In our simulations, the time step was
chosen so that the effect of thermostatting the system was not evident
in any of the dynamical correlation functions examined. This procedure
resulted in a time step of $3\rm\ fs$ for the low density runs
and $0.9\rm\ fs$ for the high density simulations of the
continuous\mbox-potential system.  The efficiency of the simulations in
the discontinuous model depends on the grid time interval used in the
root search, and here a value of $20$~fs was used for all simulations.
In general, one can use the angular velocities of the colliding
molecules to extract an average oscillation frequency of the collision
indicator function and estimate the appropriate grid time
interval for the simulation\cite{DeMichele} from this information.  In
such an approach, the angular velocities, and hence the value of
the grid interval, vary with temperature in much the same way that
the optimum time step value changes in a continuous MD simulation with
temperature.  Instead, we have chosen to use $k_{\rm B}T$ as the
energy unit in the simulation so that the average angular
velocities in the body frame have the same numerical value
regardless of the temperature.  This approach has the benefit
that the numerical value for the optimal grid time interval is
also independent of the temperature.

Figure~\ref{efficiency-ld} shows the CPU time in seconds required by
the two methods to simulate 1 ps of real time dynamics for the
low density system as a function of the number of molecules.  Note
that the the discontinuous method is faster than the MD simulation by
two orders of magnitude for all system sizes.  This result is
understandable on the basis that the dynamics in a low density system
is dominated by free molecular motion which, as discussed earlier, can
be integrated and propagated with large ``finite'' time
steps. This is to be contrasted with the ordinary MD method in which
an (ideally infinitesimally) small time step must be utilized in order to
numerically integrate the dynamics and conserve energy.  We also
observe in Fig.~\ref{efficiency-ld} that both methods scale linearly
with system size due to the use of cell lists and a cut off
interaction distance in the continuous model.

\begin{figure}
\centerline{\includegraphics[angle=0,width=0.85\columnwidth]{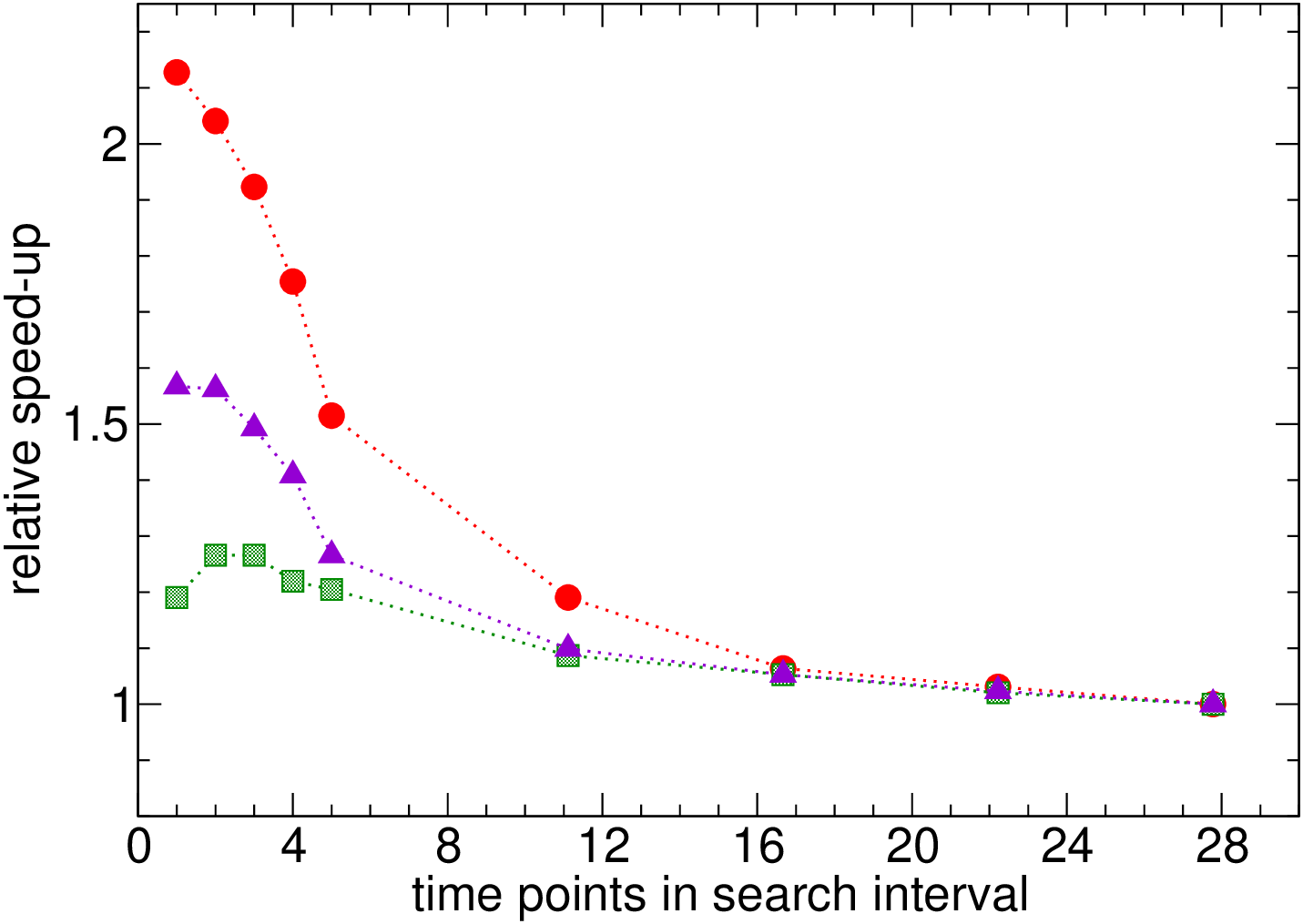}}
\caption{The relative speed\mbox-up due to the use of virtual collisions
as measured by the ratio of the CPU time for simulating liquid
methane for 1 ps without and with virtual collisions as a
function of the number of time points in the search interval.  The
circles, triangles and squares correspond to the system with 1000,
3375 and 8000 methane molecules, respectively.  }
\label{virColl}
\end{figure}

Figure~\ref{efficiency-hd} shows the CPU time needed by the DMD and MD
methods to carry out 1 picosecond of real time dynamics for the
denser system at $0.347 \rm\ g/cm^3$.  The insert shows the CPU
time ratio MD/DMD and indicates that, even at this density, the DMD
method is more than 2 times more efficient than the continuous MD
method.

One clear disadvantage of performing DMD simulations on rigid bodies
as opposed to simple hard spheres is the necessity to find collision
times using numerical root\mbox-search methods.  The evaluation of
distances between possibly interacting atoms is
computationally costly, and is often the most demanding task in a DMD
simulation of rigid systems.  To reduce the computational work of
finding collision times, it is convenient to truncate the numerical
search of collisions that extend beyond a certain maximum time
interval, which will be called the {\em search interval},
since events far in the future are not likely to be executed.  To
implement a truncation scheme, collisions found in 
the search interval are scheduled as such, while a truncated search is
scheduled as a \emph{virtual collision} (see
Ref.~\onlinecite{preceeding}). If the molecules for which the search
for a collision event was truncated do not have an event over the
previously searched interval, a virtual collision event is executed
for the pair in which the root search is continued from where it was
previously stopped.  In this way, the numerical effort is primarily
focused on the most likely events (i.e. the ones happening within the
search interval) and the number of grid evaluations in which the
Cartesian positions and the distances between all atoms in the pair of
molecules are calculated, is diminished.  Even though the truncation of
the numerical search for collision events reduces the computational
load of this type of event, the introduction of new events increases
the size and complexity of the binary tree used to manage events.  As
a result, the scheduling of new events, which typically scales as
$N\log N$, where $N$ is the number of events in the tree, becomes more
demanding.  Since the computational effort associated with the data
management increases while the effort of finding collisions decreases,
the optimal size of the time interval for the virtual collisions is a
function of the size of the tree, and hence the overall size of the
system.  If the search interval is very large, very few virtual
collisions are scheduled and the efficiency of the simulation
converges to that of a simulation that does not include virtual
collision events.

\vfill

Fig.~\ref{virColl} shows the relative gain in efficiency due to the
use of virtual collision events for the systems consisting of 1000,
3375 and 8000 molecules as a function of the maximum number of
evaluations of the collision function in the search interval.
Clearly, the use of virtual collision increases as this ratio is
decreased.  For the system with 1000 molecules, an intensive use of
virtual collisions (i.e. one or two grid evaluations) doubles the
computational efficiency with respect to a simulation not utilizing
these events. Smaller systems (not shown) exhibit the same trend and
have similar relative efficiencies.  For a very large system, however,
the gain in efficiency is smaller due to the fact that a large binary
tree slows down the processing of events. Indeed, for the system with
8000 molecules, it is more efficient to limit the number of virtual
collisions scheduled by extending the search up to two or three grid
evaluations instead of performing a single grid evaluation per root
search.  The effect of the increased load in data management can be
clearly seen as the system size increases.  Nonetheless, the
scheduling of truncated root searches leads to significant
improvements in simulation efficiency for all system sizes examined
where the numerical search for collision times, as opposed to data
management, represent the bulk of the computational work in the
simulation.

\section{Benzene}
\label{Benzene}

\subsection{Molecular model}

The methane system considered in the previous section has particularly
simple rotational dynamics due to the high degree of symmetry of the
methane molecule.  However\, the formalism established in
Ref.~\onlinecite{preceeding} allows simulations of rigid molecules
with arbitrary mass distribution to be performed.  A simple example of
slightly more complicated rotational dynamics is that observed in
symmetric top molecules, such as benzene, in which two of the
principal moments are equal and the third is distinct.

For a system of rigid benzene molecules, there are a total of 13
relevant sites in every molecule; the center of mass, six carbon
sites, and six proton sites, where all atoms lie in a single plane
with the carbon sites forming a closed ring.  However\, using a
\emph{united atom} approach, in which every carbon\mbox-hydrogen pair is
represented as a unique site, the number of relevant sites in every
molecule is reduced to 7. If the radial distance from the center of
mass to each united ``carbon'' atom in the plane of the molecule
(taken to be the $XY$ plane) is $d$ and the mass of the united atoms
is set to $m_{\rm CH} = 13 m_p$, then one finds that the principal
moments of inertia are $I_{1} = I_{2} = 3 m_{\rm CH}
d^2= 39 m_p d^2$ and $I_{3} = 2I_{1}$.

Typically, the intermolecular interactions between the united atoms in
benzene are assumed to be of Lennard\mbox-Jones form, leading to an
interaction energy that depends strongly on the relative orientations
of the interacting molecules.  The simplest possible
discontinuous\mbox-potential model possessing attractive but directional
interactions consists of a combination of a spherically symmetric
attractive well about each center of mass combined with purely repulsive
interactions between united carbon atoms.  We therefore consider a
united atom model of benzene in which the center of mass site, or site
$1$, is defined to have a repulsive hard\mbox-wall interaction with any
other center of mass site, at the interaction distance $d_{11}$
while the united carbon sites repel one another at distance
$d_{\rm CC}$. 
The attractive interaction potential between the centers of mass
of two molecules is defined as
\begin{equation}
V_{11}^{\rm attr}\left( r_{11}^{ab}  \right) =
\begin{cases}
-V_{SW1} & \text{if } r_{11}^{ab} \leq d_{SW1} \\
-V_{SW2} & \text{if } d_{SW1} < r_{11}^{ab} \leq d_{SW2} \\
\ \ \ 0   & \text{if } r_{11}^{ab} > d_{SW2},
\end{cases}
\label{squareWell2}
\end{equation}
where $r_{11}^{ab}$ is the distance  between the center of mass of any
two molecules  $a$ and $b$.   This potential contains two attractive
square\mbox-well regions and  four parameters:  two square\mbox-well distances
$d_{SW1}$ and $d_{SW2}$, and two square\mbox-well depths $V_{SW1}$
and  $V_{SW2}$.

A DMD simulation with this model can be implemented in the same way as
for the methane model in Sec.~II, with a slightly more complex
propagation matrix $\mathbf P(t)$ (see Eq.~(\ref{Asolution})).  For
the symmetric top system, $\mathbf P(t)$ can be written as the product
of two rotation matrices\cite{preceeding},
\begin{equation}
  \mathbf P(t) =
  \mathbf U(-\omega_p t\hat{\bm z}) \cdot \mathbf
  U\left(-\frac{\tilde{\bm L}_0}{I_{1}} \, t \right),
\label{Psymmtop}
\end{equation}
where $\omega_p =
\left(1-I_{3}/I_{1}\right)\tilde{\omega}_{z}(0)$ is called
the precession frequency, and $\tilde{\bm L}_0 = \tilde{\bm I} \cdot
\tilde{\bm \omega}(0)$ is the initial angular momentum vector in the
body frame.  Unlike the case of the spherical rotor, the $x$ and $y$
components of the angular velocity in the body frame are
time\mbox-dependent, and the moment of inertia tensor in the lab frame is
no longer constant and diagonal. Thus, not only is the rotational
motion more complicated, but the processing of collisions involves
explicitly inverting the moment of inertia tensor to evaluate the
change in angular velocities via Eq.~(\ref{changeOmega}).

For the purposes of comparison, a continuous\mbox-potential
united\mbox-atom model of benzene, in which the united\mbox-atoms
interact via intermolecular Lennard\mbox-Jones interactions defined in
Eq.~(\ref{softmodel}), was considered.  Note that this model does not
posses any explicit interactions between centers of mass and that the
site-site interactions lead to an effective orientational dependence
in the interaction between molecules.

\begin{figure}
\includegraphics[width=0.85\columnwidth,angle=0,clip]{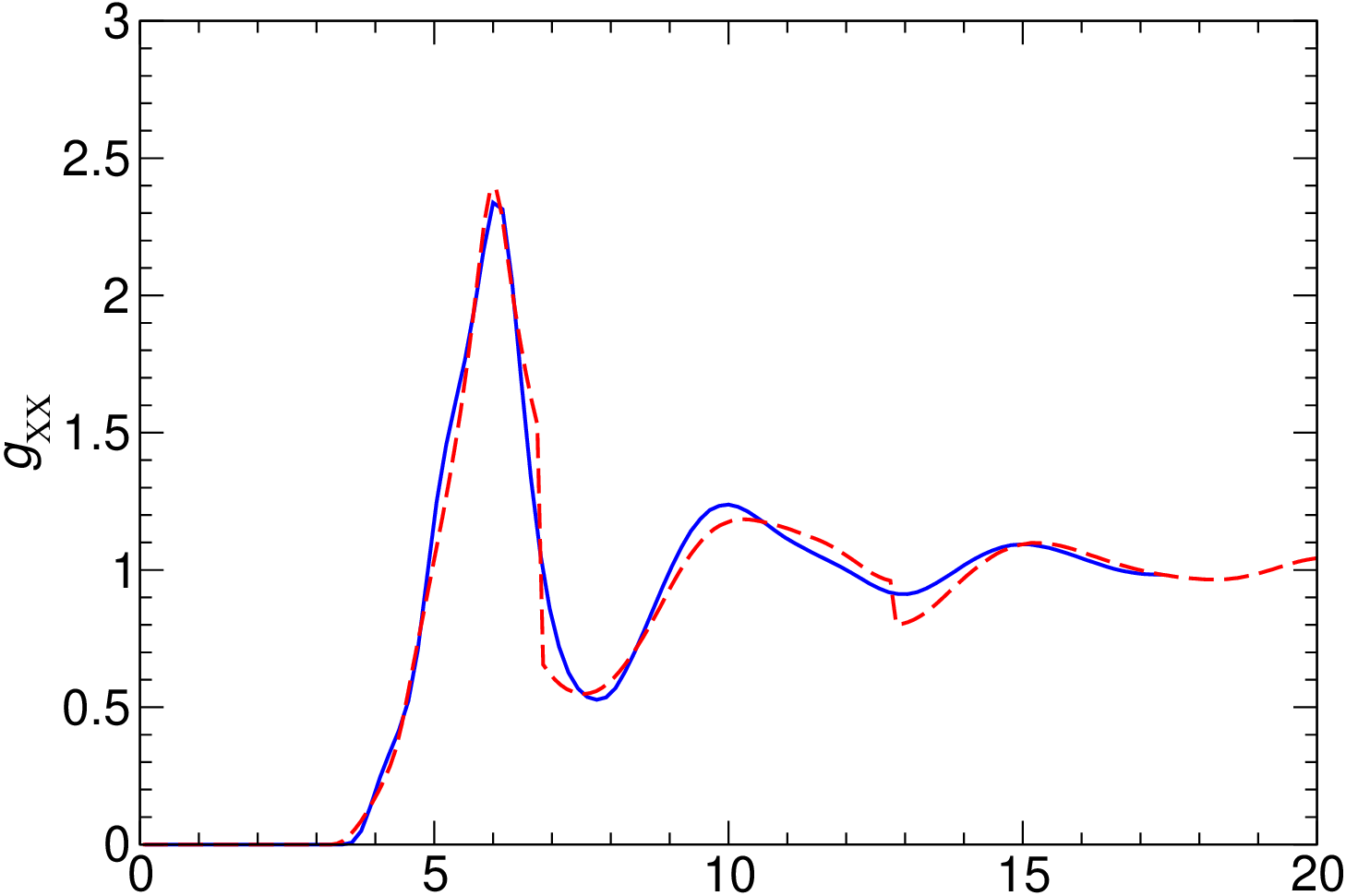}
\includegraphics[width=0.85\columnwidth,angle=0,clip]{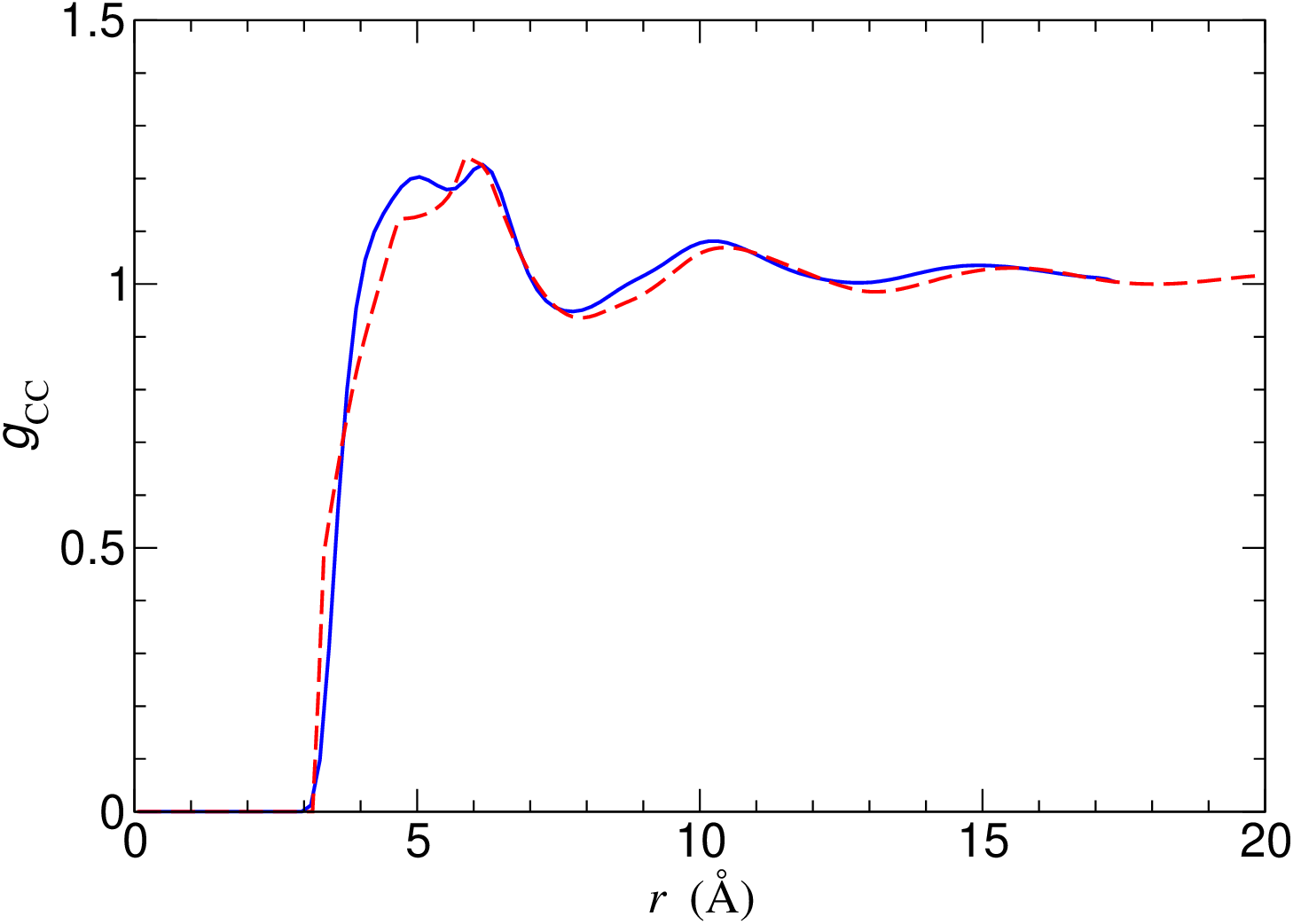}
\caption{From top to bottom, center of mass\mbox-center of mass (XX)
and carbon\mbox-carbon RDFs for liquid benzene.  The continuous lines
represent the results for continuous MD model and the dashed lines
represent those for the DMD model (see Sec.~\ref{details} for
details).}
\label{radial-Benzene}
\end{figure}

\subsection{Simulation details}
\label{details}

Simulations were performed for both models in a box with 512 molecules
at a density of $0.870$~$\rm g/cm^3$ with initial conditions drawn
from an iso-kinetic ensemble at a temperature of $298$~K ($k_{\rm
B}T=2.478$~kJ/mol).  The distance $d$ between carbon sites (or between
the center of mass and every carbon site) was set to 1.48~\AA~in both
models of rigid benzene.

The parameters in the DMD model used to obtain the results presented
below are: $d_{CC}=3.25$~\AA, $d_{11}=2.0$~\AA, $d_{SW1}=6.8$~\AA,
$d_{SW2}=12.8$~\AA, $V_{SW1}=1.982$~kJ/mol and $V_{SW2}=0.446$~kJ/mol.
The numerical search for collision events were carried out using a
time grid of $19$~fs and with only one time point in the search
interval (i.e. extensive use of virtual collisions).  The simulation
was equilibrated with frequent rescaling of velocities to arrive at an
initial condition drawn from an iso-kinetic ensemble, and later run
with rescaling of the velocities only every picosecond.

The Lennard\mbox-Jones parameters used for the continuous model
simulations are $\epsilon_{ij}=0.457$~kJ/mol and
$R_{min}^{ij}=3.695$~\AA.  The simulations were carried out in a
standard manner using a predictor\mbox-corrector integrator on the
translational and the orientational (quaternion) parameters. A time
step of $1$ \rm{fs} and a cut off distance of $8.0$~\AA~ for the
truncation of the intermolecular interactions were used. With these
parameters, the DMD method was 3 times more efficient computationally
(measured as the CPU time needed for the simulation of one picosecond
of real\mbox-time dynamics) than the MD simulation.  The continuous model
was simulated at constant temperature with a thermostat to correct for
the numerical energy drift.

\subsection{Static properties}

The average intermolecular energy values obtained for the DMD and the
continuous model were 
$-23.558\pm0.005$~kJ/mol
and $-25.4\pm0.1$~kJ/mol,
respectively. 
The
pressure values were $930\pm25$~bar and $360\pm60$~bar for the step
potential model and the MD simulation, respectively.

Fig.~\ref{radial-Benzene} shows the center of mass-center of mass and
carbon\mbox-carbon radial distribution functions obtained for the DMD
and continuous models of liquid benzene.  The results agree with one
another and with previous molecular dynamics studies\cite{liqbenzene}
of liquid benzene at similar densities and temperatures. It is clear
that the DMD model reproduces most of the structural details found
with the continuous model in both radial distribution functions.  In
Fig.~\ref{radial-Benzene}, the DMD result for the center of mass
radial distribution function shows two sharp decreases at distances of
about $6.8$ \AA~and $13.8$ \AA, which coincide naturally with the
values of the square well attractive potential distances chosen for
the model. Indeed, these distances were chosen such that the DMD model
can match the continuous radial distribution functions.

\subsection{Dynamical properties}

\begin{figure}
\includegraphics[angle=0,width=0.85\columnwidth]{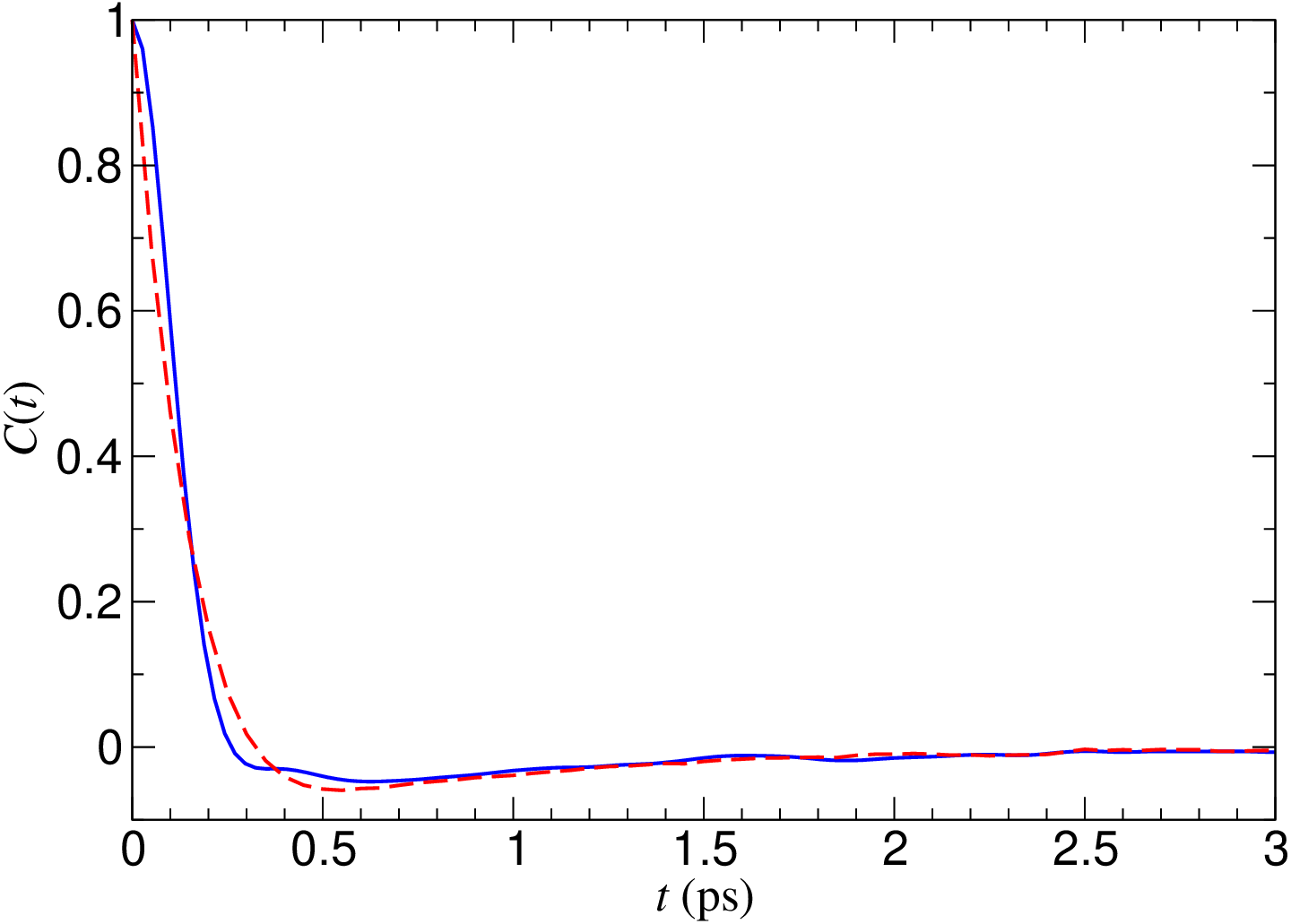}
\caption{Normalized velocity time autocorrelation function for
liquid benzene.  The continuous line corresponds to the Lennard\mbox-Jones continuous
potential
and the dashed line corresponds to the DMD result. }
\label{TimeCorrelationFunction}
\end{figure}

The normalized center of mass velocity time autocorrelation functions
obtained for the DMD and MD models are shown in
Fig.~\ref{TimeCorrelationFunction}.  In spite of the differences
between the models, the agreement between the correlation functions is
very good.  The values of the diffusion coefficients, obtained from
integrating the functions in this figure, are
$(1.42\pm0.04)$~$\cdot10^{-5}\rm cm^2/s$ and
$(1.62\pm0.04)$~$\cdot10^{-5}\rm cm^2/s$ for the DMD and MD
simulations, respectively.  The values agree well, although both
are smaller than the experimental value of $2.27$~$\cdot10^{-5}\rm
cm^2/s$\cite{Diffbenzene} reported at this temperature.

\begin{figure}
\includegraphics[angle=0,width=0.85\columnwidth]{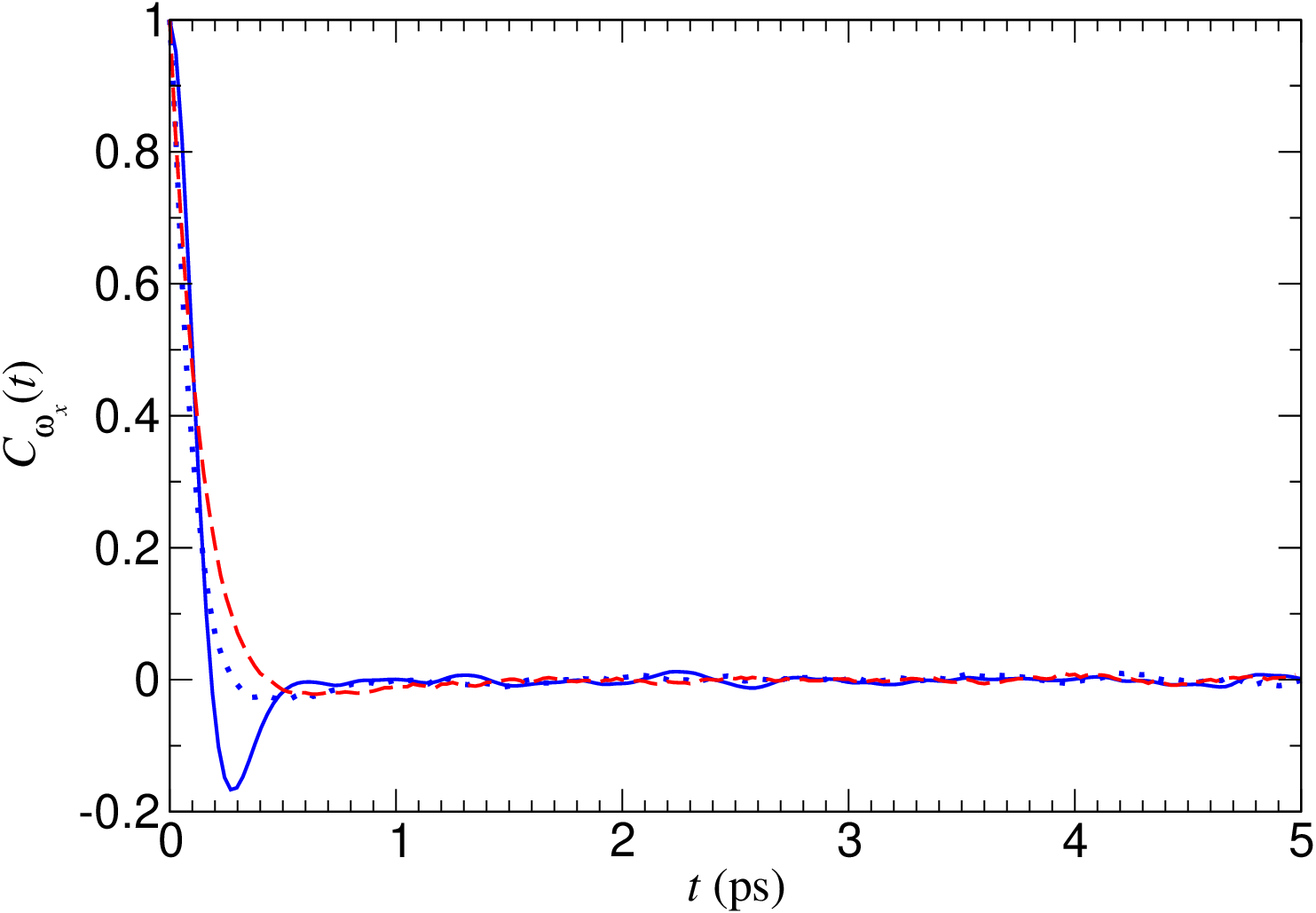}
\includegraphics[angle=0,width=0.85\columnwidth]{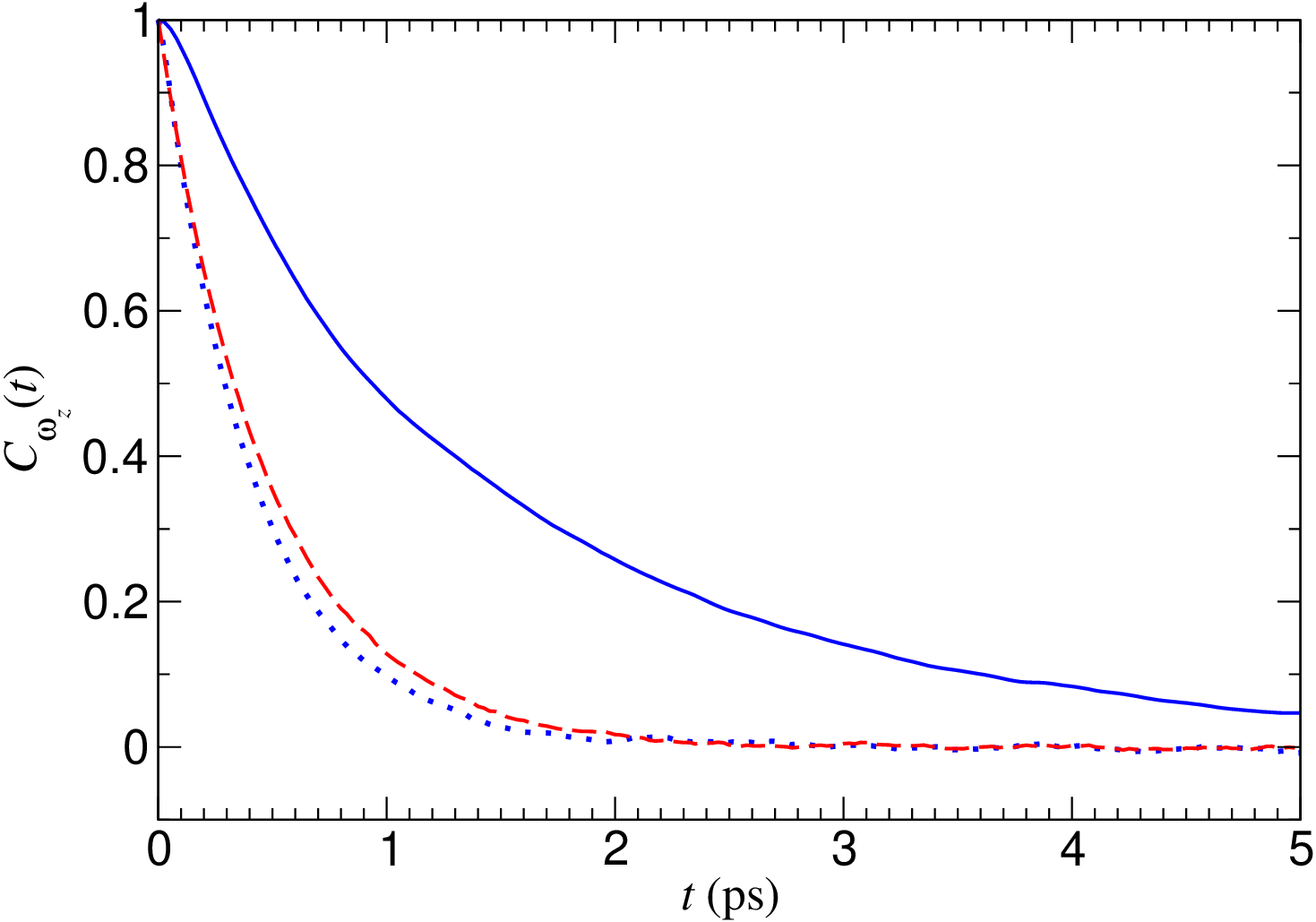}
\caption{Time correlation functions of the $x$ and $z$ components of
the angular velocity vector in liquid benzene.  The continuous line
corresponds to the Lennard\mbox-Jones potential and the dashed line
corresponds to the DMD result. The dotted lines are the results
obtained with the hyperbolic tangent potential defined by
Eq.~(\ref{hypTan}).}
\label{angularTimeCorrelationFunction}
\end{figure}

The rotational motion in liquid benzene is more complex than the
motion in liquid methane, and in other spherical rotor systems, due to
the asymmetry of the benzene molecule.  The asymmetry leads to
non\mbox-trivial orientational components of the static structure in which
the planes of nearby benzene molecules tend to be orthogonal to one
another.  The asymmetry also leads to interesting correlations in
rotational motion that can be examined by looking at normalized
autocorrelation functions $C_{\omega_k} (t)$ of components of the angular
velocity in the body frame,
\begin{equation}
C_{\omega_k}(t)   =    \langle   \tilde{\omega}_k(0)   \tilde{\omega}_k(t)\rangle   /
\langle\tilde{\omega}_k^2(0)\rangle ,
\end{equation}
where the $\tilde{\omega}_k$ are the $k=x,y,z$ components of the angular
velocity vector in body frame. In
Fig.~\ref{angularTimeCorrelationFunction}, the time autocorrelation
functions of the $x$ and $z$ components of the angular velocity vector
in the body frame are plotted versus time, where $x$ and $y$ are the
principal axes in the plane of the molecule and $z$ is the principal
axis ($C_6$\mbox-axis) orthogonal to the plane of the molecule.  Note that,
due to symmetry, $C_{\omega_x}(t)=C_{\omega_y}(t)$.

As can be seen in Fig.~\ref{angularTimeCorrelationFunction}, the $x$
component of the angular velocity time correlation function for the
continuous model exhibits a well defined minimum that is not
reproduced in the DMD model.  Angular correlations of these components
are short\mbox-lived in the dense system with a lifetime of less than a
picosecond, due to the strong confining effect of molecular cages in
the liquid.  The minimum in $C_{\omega_x}(t)$ calculated in the
continuous potential model reflects an anti\mbox-correlation in the angular
velocities due to molecules rebounding off of nearest neighbors and is
reminiscent of anti\mbox-correlations observed in velocity autocorrelation
functions.  Interestingly, this phenomenon is not observed in the DMD
simulation. 

In contrast, the autocorrelation function of the $z$ component of the
angular velocity has a much longer lifetime in both models, although
$C_{\omega_z}(t)$ decays much too quickly in the DMD simulation.  The
longer lifetime arises due to the relatively free motion of rotations
about the $z$\mbox-axis of a molecule, corresponding to a planar
spinning motion.

\begin{figure}
\includegraphics[angle=0,width=0.85\columnwidth]{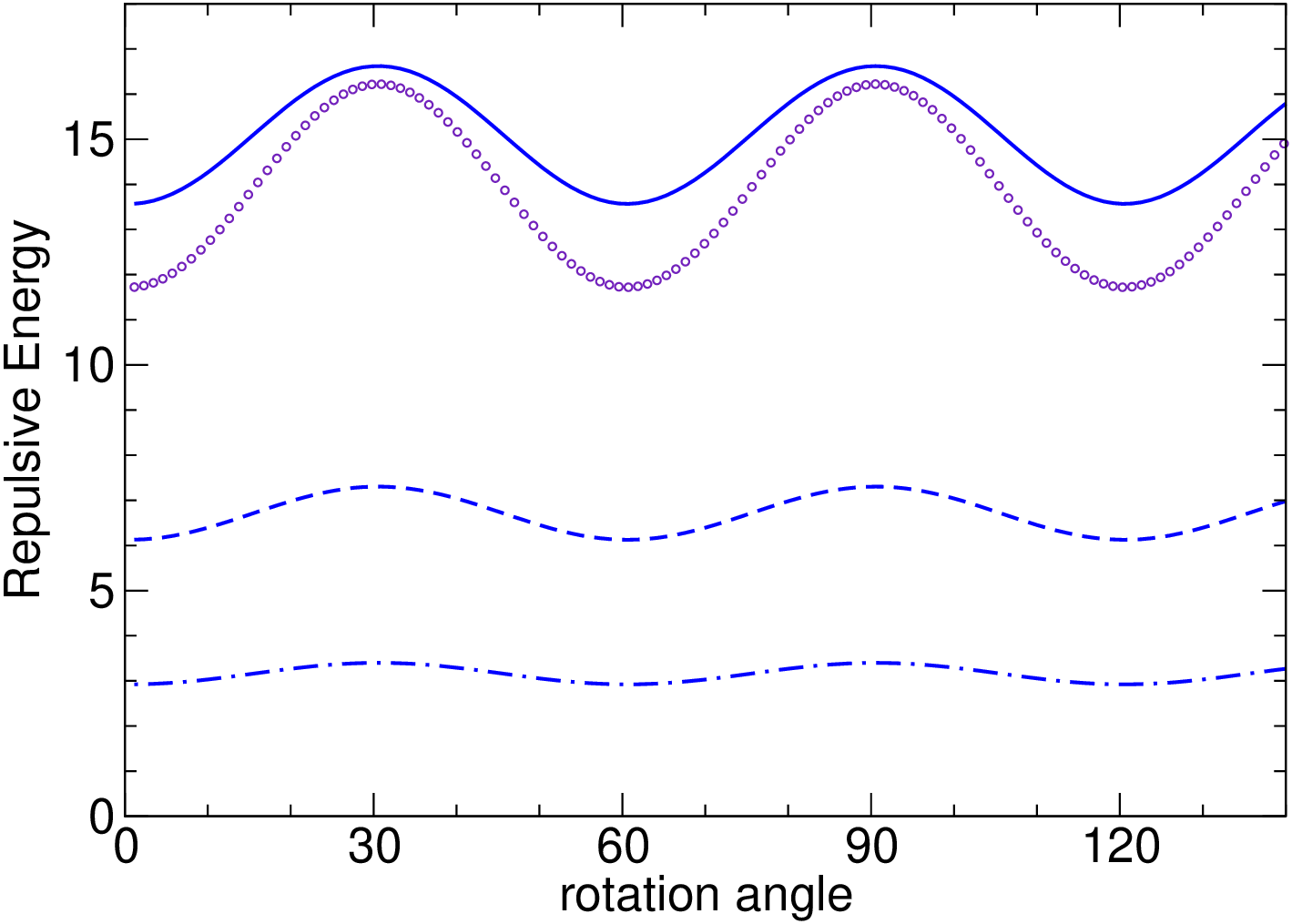}
\caption{Repulsive potential energy (in units of $k_BT$) for the
continuous potential model between two benzene molecules as a
function of their relative orientation measured by the rotation angle
(in degrees) around the $z$\mbox-axis of one the molecules. The
continuous, dashed, and dot\mbox-dashed lines correspond to coplanar
benzene molecules at center\mbox-of\mbox-mass distances $r_{11}$ of
$6.0$~\AA, $6.25$~\AA, and $6.4$~\AA, respectively. The circles
correspond to molecules in orthogonal planes with $r_{11}=6.0$~\AA.}
\label{repulsiveEnergy}
\end{figure}

In order to understand the origin of some of the deficiencies of the
discontinuous model, it is instructive to examine the interaction
energy between two benzene molecules, lying in the same plane, as
one molecule is rotated relative to the other around the $z$\mbox-axis for
both models.   This interaction energy for the
continuous potential model is plotted in Fig.~\ref{repulsiveEnergy}.
In the discontinuous model, the rotational motion is completely free
unless the centers of mass of the molecules are close enough
($r_{11}<6.25$ ~\AA) so that a hard\mbox-core collision takes place.
When the hard\mbox-core collision occurs, the $z$\mbox-component of
the angular velocity in the body frame changes direction and the
rotation effectively reverses direction.  In contrast, the repulsive
energy in the continuous model increases up to a maximum where the
distance between united\mbox-carbon atoms on the two molecules is
minimized.  The repulsive energy drops as the molecule rotates past
this point, so that there is a ``hindered'' rotational interaction for
the rotation of one molecule past another.  Such oscillatory
interactions, clearly absent in the DMD model, are responsible for the
qualitative differences observed in
Fig.~\ref{angularTimeCorrelationFunction}.  Note that the attractive
term of the Lennard\mbox-Jones interaction can be effectively
accounted for via center of mass interactions, and that its inclusion
in this analysis will only further enhance the oscillatory structure
of the interaction.  An interaction analogous to the one
shown in Fig.~\ref{repulsiveEnergy} can be included in the DMD model
by adding finite repulsive shoulder interactions between united
carbon atom sites.

Alternatively, the role of the hindered rotational interactions on the
angular correlations can be isolated by systematic modifications of
the continuous interaction potential to approach the impulsive
potential limit and, therefore, the DMD model.  For example, instead of
using a Lennard\mbox-Jones interaction, repulsions between united carbon
atoms can be modeled with a potential of the form,
\begin{equation}
V^{\text{rep}}_{ij}\left( r_{ij}^{ab} \right) =
\alpha_1 \tanh \left\lbrack \alpha_2 (d_{ij} - r_{ij}^{ab} ) \right\rbrack.
\label{hypTan}
\end{equation}
The parameters $\alpha_1$ and $\alpha_2$ can be independently assigned
to control the height and the width of the repulsive potential,
respectively. The value of $\alpha_1$ can be set large enough that the
probability of any molecule going through the barrier is virtually
zero.  The hard\mbox-wall limit in which the interaction range approaches
zero and interactions become instantaneous is formally obtained in the
limit of infinite $\alpha_2$.  Of course, such a potential cannot be
integrated numerically, and comparisons of dynamics using numerical
integration schemes and dynamics in the impulsive limit must use a
large, but finite value of $\alpha_2$.

\begin{figure}
\includegraphics[angle=0,width=0.85\columnwidth]{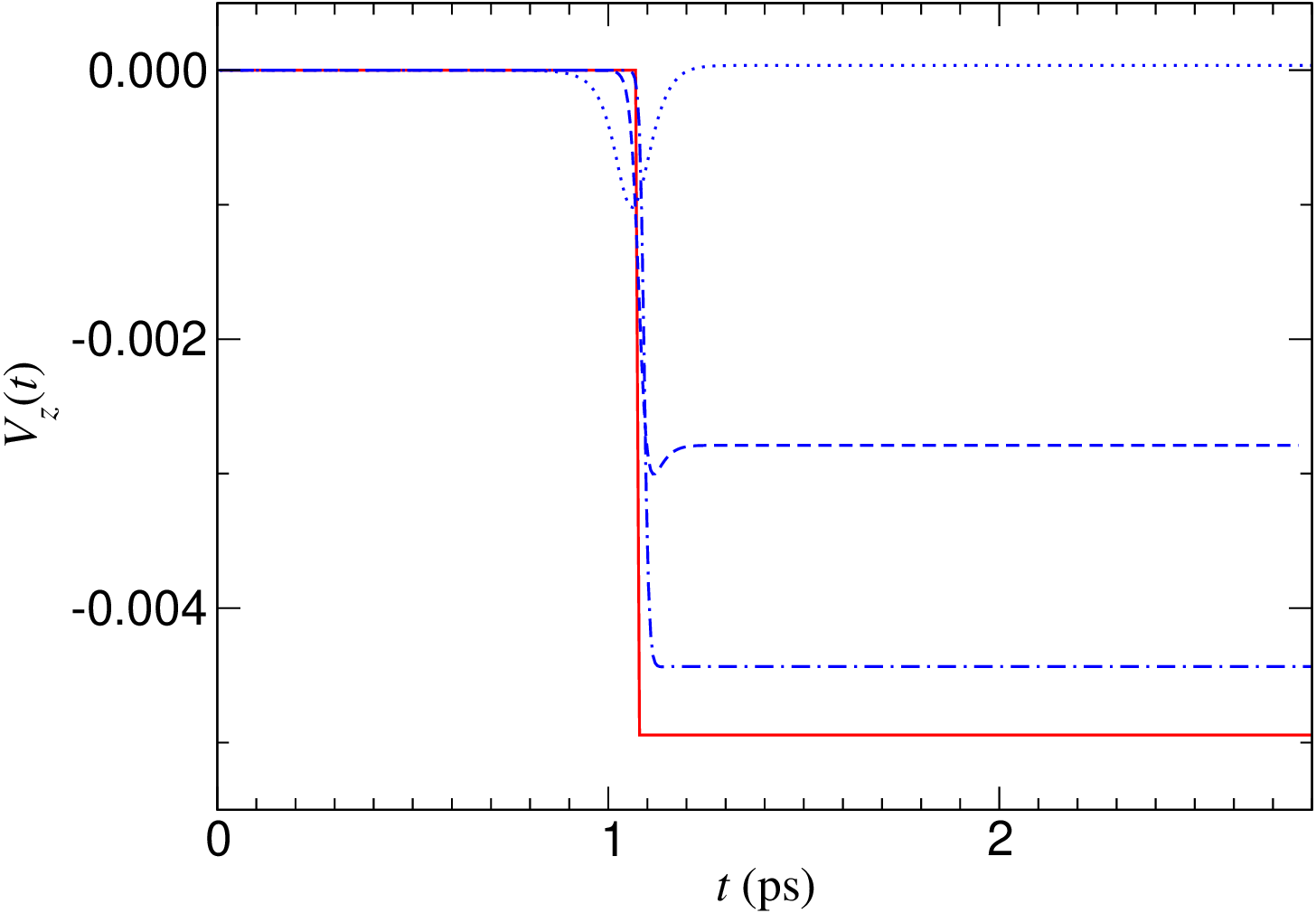}
\includegraphics[angle=0,width=0.85\columnwidth]{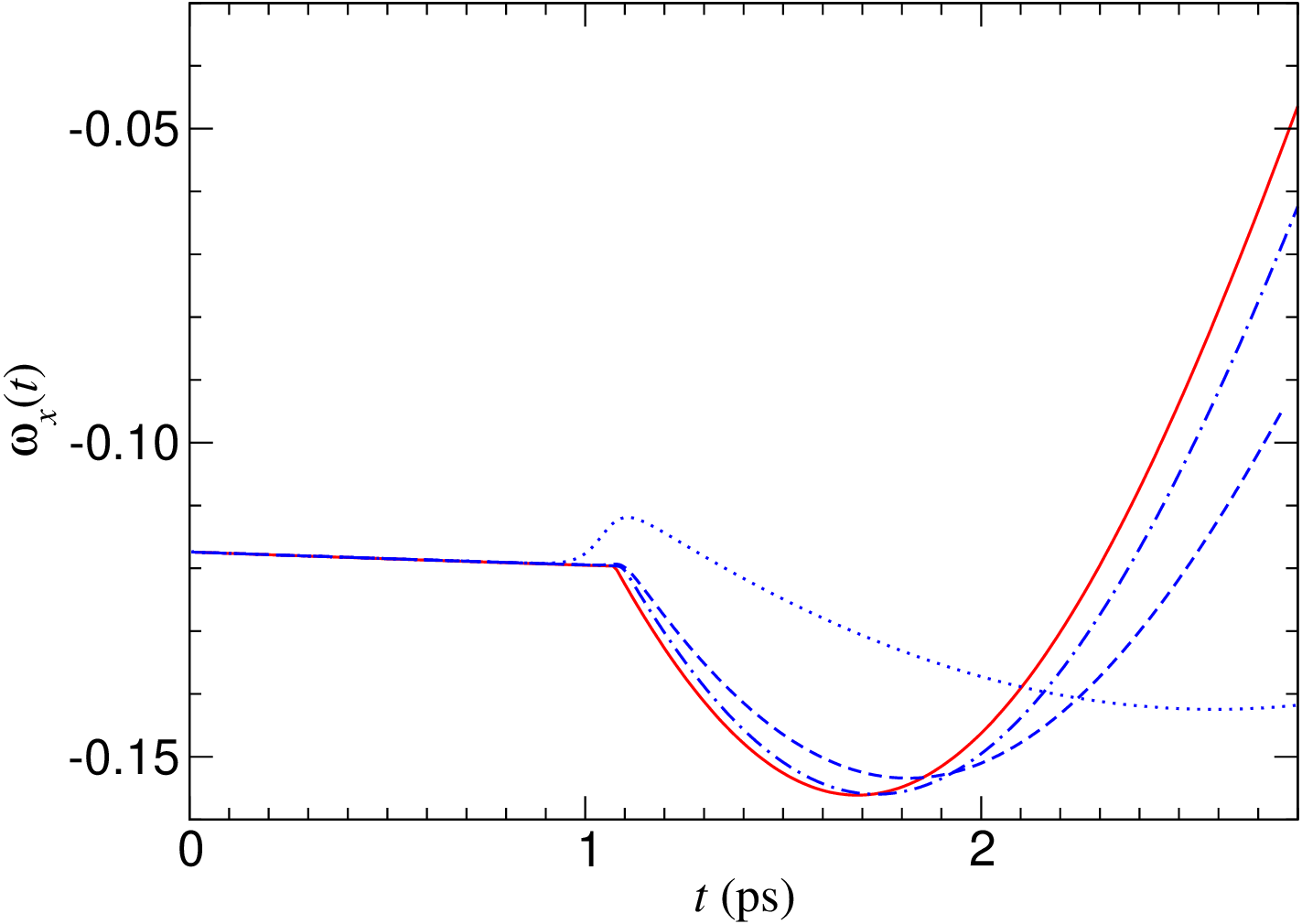}
\caption{Time evolution of the $z$ component of the velocity vector
(top panel) and the $x$ component of the angular velocity vector
(bottom panel) of a benzene molecule involved in a collision
with another benzene molecule. The solid lines correspond to the
DMD simulation. The dot, dashed and dot\mbox-dashed lines correspond to the
model interactions given by Eq.~(\ref{hypTan}) with $\alpha_2$
values of 5, 10 and 20, respectively.}
\label{Converegence1}
\end{figure}

As an example, consider the trajectories of two benzene molecules
before and after their mutual interaction through the potential
defined by Eq.~(\ref{hypTan}).  In Fig.~\ref{Converegence1}, the
dynamical evolution of separate components of the linear and angular
velocity vectors of one of the molecules involved in the collision is
shown as a function of time for the DMD model.  The trajectories
in the impulsive limit are also included in this figure for
comparison.  Since the center of mass velocity vector, its components,
and the $z$ component of the angular velocity vector are constants of
the motion, their behaviour is analogous to the one shown in the
top panel of Fig.~\ref{Converegence1}.  It is clear that the
interaction time goes to zero and the final $v_z$ value approaches the
DMD result as the value of $\alpha_2$ increases. The $x$ and $y$
components of the angular velocity vector, on the other hand, follow
an oscillatory dynamics and their time evolution after the interaction
can be quite different than that observed in the impulsive limit, as
can be seen in the bottom panel of this figure. Once again, one
observes that the dynamics generated by the continuous potential
approaches the DMD dynamics as the value of $\alpha_2$ increases.

The time\mbox-correlation results obtained with continuous MD
simulations of a system composed of 512 benzene molecules interacting
via Eq.~(\ref{hypTan}), at the same density and temperature as above,
are also shown in Fig.~\ref{angularTimeCorrelationFunction}.  These
results are obtained with the values $\alpha_1=40$ and $\alpha_2=20$
for the carbon\mbox-carbon repulsion. Not surprisingly, the new time
correlation functions agree much better with results obtained with the
DMD model than those of the Lennard\mbox-Jones model.  This agreement
confirms that the discrepancies observed in the angular velocity
correlation functions in the discontinuous and continuous models are a
consequence of the overly simplistic interaction potential between
united carbon sites.

\begin{figure}
\includegraphics[angle=0,width=0.85\columnwidth]{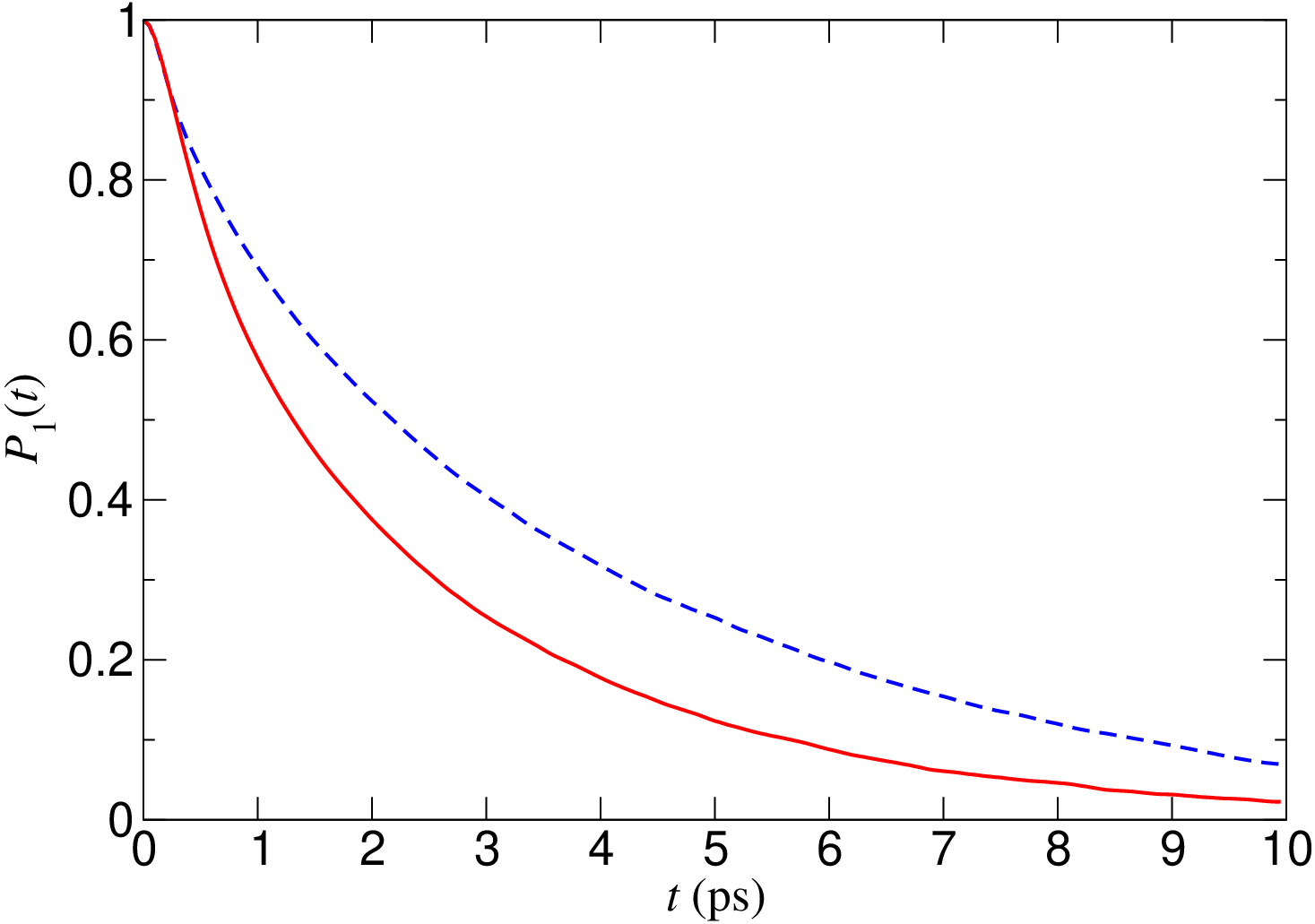}
\includegraphics[angle=0,width=0.85\columnwidth]{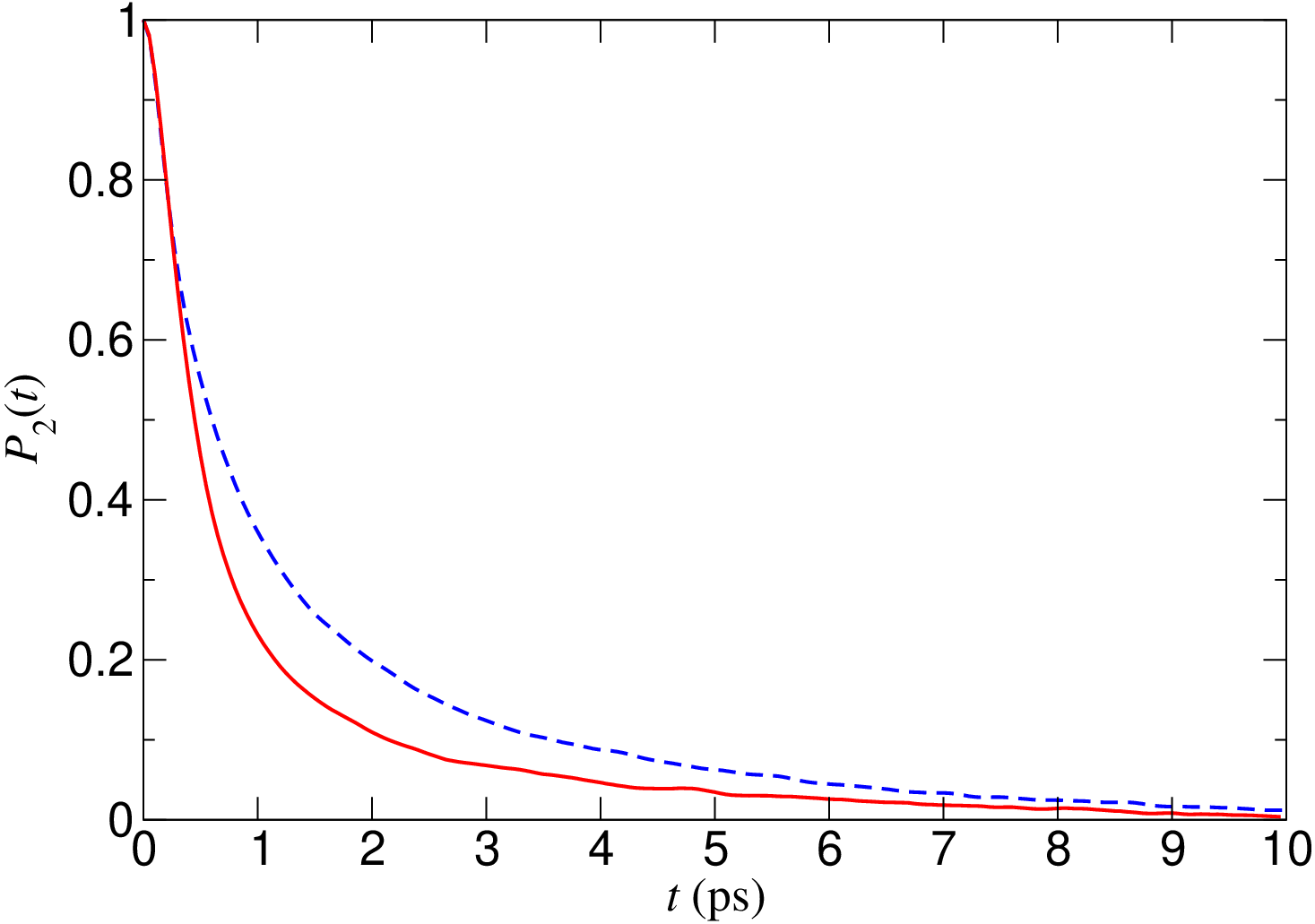}
\caption{Orientational time correlation functions $P_1(t)$ and
$P_2(t)$ in liquid benzene.  The continuous line corresponds to the
Lennard-Jones potential and the dashed line corresponds to the DMD
result.}
\label{P1andP2TimeCorrelationFunction}
\end{figure}

In addition to correlation functions of the angular velocities, one
can also examine how long individual benzene molecules retain their
orientation in a dense liquid.  One common measure of orientational
correlation involves computing ensemble averages of low order Legendre
polynomials of the cosine of the angle between an orientational vector
of a molecule, such as the $C_6$ axis (or $z$\mbox-axis), at two times.  In
Fig.~\ref{P1andP2TimeCorrelationFunction}, the $P_1(t)$ and $P_2(t)$
orientational time correlation functions, defined by
\begin{equation}
P_1(t) = \langle \hat{\bm z}(0)\cdot \hat{\bm z}(t) \rangle  
\label{P1}
\end{equation}
and
\begin{equation}
P_2(t) = \left\langle \frac{1}{2} \left\lbrace 3 \lbrack \hat{\bm z}(0)\cdot \hat{\bm z}(t) \rbrack^2 
-1 \right\rbrace \right\rangle
\label{P2}
\end{equation}
for the DMD and Lennard\mbox-Jones models are shown.  The overall shape of
these functions is in agreement with previous molecular dynamics
simulations of liquid benzene\cite{liqbenzene1}.  Although there is
qualitative agreement in orientational correlations in the continuous
and discontinuous models, again it is evident that orientational
correlations typically decay more quickly in the discontinuous model
than in the continuous model.  It is likely the longer\mbox-lived
orientational correlations in the continuous case arise due to the
hindered rotational interactions that are poorly modeled in the simple
discontinuous system.

\section{Conclusions}
\label{Conclusions}

In this paper DMD simulations of two rigid molecular systems, which
differ in the symmetry of their mass distribution, were presented.
The simulation methodology utilized is based on exact solutions of the
free motion of rigid bodies in combination with collision rules
derived from conservation laws.  In Sec.~\ref{Methane}, a
discontinuous model of rigid methane was constructed in which methane
molecules attract one another via a center of mass square\mbox-well
potential, while the constituent atoms of different molecules repel
each other through hard\mbox-core repulsions.  DMD simulations of the
spherical rotor system were carried out in low and high density
regimes, and excellent agreement with standard molecular dynamics
simulations was observed for all structural quantities.  It was
furthermore found that dynamical quantities, such as the
self\mbox-diffusion coefficient, also agree well with those of
standard MD simulations.

In addition, the efficiency of DMD simulations relative to standard
continuous\mbox-potential simulation methods of the rigid
methane system was examined in order to assess the role of certain
parameters such as the grid search size and the truncation interval
for the root search in the DMD simulations.  It was found that the DMD
simulations were between 2.5 and 100 times more efficient than
standard continuous MD simulations at densities of $0.347 \rm\ g/cm^3$
and $3.47 \cdot 10^{-3} \rm\ g/cm^3$, respectively.  Furthermore, it
was found that it is generally most efficient to schedule
virtual\mbox-collision events (i.e.  truncated root searches) as
frequently as possible for small systems and slightly less frequently
for larger systems.  The use of virtual\mbox-collision events typically
leads to improvements in simulation efficiency of 25\% for large
systems and up to 250\% for small systems.

In Sec.~\ref{Benzene}, a united\mbox-atom model with
discontinuous potentials for rigid benzene was presented in which
benzene molecules attract one another via center of mass interactions,
while united carbon atoms repel one another via hard\mbox-core
interactions.  DMD simulations of the rigid benzene system at a liquid
density were carried out using the simulation methodology appropriate
for a symmetric\mbox-top rigid body.  To the best of our knowledge,
this is the first event driven simulation of a non\mbox-spherical
rigid body, other than a linear dumbbell, where the dynamics and
collision rules are properly taken into account without approximation
of the dynamics or collision rules.  Static and dynamical quantities
from the DMD simulation were compared with those obtained from a
continuous potential benzene model, in which the united carbon atoms
interact via a standard Lennard\mbox-Jones potential and found to be
in good agreement.  The quantitative discrepancies in
dynamical correlations of angular velocities and orientational
directors between the models are readily explained in terms of the
crude level of description in the discontinuous model of the hindered
rotational motion of one benzene molecule near
another. Nonetheless, one can argue that the relevant physics is
already present in the crude description of the interactions in the
DMD model, such as long-lived orientational correlations and
correlation times of $\omega_z$ which are significantly larger than
those of $\omega_x$ and $\omega_y$.  If one is interested in more
detailed features of the dynamics, it is always possible to add more
discontinuities in the interactions, at the expense of computational
efficiency. In a sense, this is contrary to the spirit of the DMD
method, which is aimed at efficiently simulating the main physical
characteristics of a system.

The major purpose of this work is to establish that event\mbox-driven
simulations of rigid molecular systems can be carried out with
impressive gains in efficiency over standard molecular dynamics
methods.  Of course, the magnitude of the gain in relative efficiency
depends on a number of factors, such as the level of detail in
constructing a discontinuous potential model of the system, the system
size, and physical parameters such as the density.  
Nonetheless, our
results indicate that the gain in efficiency through the use of crude
discontinuous potential models of condensed phase systems may allow
larger systems or longer time scales to be investigated than those
currently accessible with standard simulation methods.

\section{Acknowledgments}
S.B.O.~gratefully acknowledges Evan O'Connor for useful discussions.
This work was supported by grants from the National Sciences and Engineering
Research Council of Canada (NSERC).

\end{document}